\documentclass[aps,prl,twocolumn]{revtex4}

\usepackage{epsfig, amsmath, amsfonts, amssymb, bm, graphicx,color}
\usepackage{braket}

\begin{document}

\DeclareGraphicsExtensions{.eps,.EPS,.pdf,.PDF}

\title{Exploring out-of-equilibrium quantum magnetism and thermalization in a spin-3 many-body dipolar lattice system}

\author{S. Lepoutre$^{1,2}$, J. Schachenmayer$^{3}$, L. Gabardos$^{1,2}$, B. Zhu$^{4,5}$,  B. Naylor$^{1,2}$, E. Mar\'echal$^{1,2}$, O. Gorceix$^{1,2}$, A. M. Rey$^{4}$, L. Vernac$^{1,2}$, B. Laburthe-Tolra$^{1,2}$}

\affiliation{$^{1}$\ Universit\'e Paris 13, Sorbonne Paris Cit\'e, Laboratoire de Physique des Lasers, F-93430, Villetaneuse, France\\
$^{2}$\ CNRS, UMR 7538, LPL, F-93430, Villetaneuse, France\\
$^{3}$\ CNRS, UMR 7504, IPCMS; UMR 7006, ISIS; and Universit\'e de Strasbourg, Strasbourg, France\\
$^{4}$\ JILA, NIST and Department of Physics, University of Colorado, Boulder, USA\\
$^{5}$\ ITAMP, Harvard-Smithsonian Center for Astrophysics, Cambridge, MA 02138, USA}

\begin{abstract}
Understanding quantum thermalization through entanglement build-up in isolated quantum systems addresses fundamental questions on how unitary dynamics connects to statistical physics.  Here, we  study the spin  dynamics and approach towards local thermal equilibrium of a macroscopic ensemble of  $S=3$ spins   prepared in a pure coherent spin state, tilted compared to the magnetic field, under the effect of magnetic  dipole-dipole interactions.
The experiment uses a unit filled array of $\approx 10^4$ chromium atoms  in a three dimensional optical lattice, realizing the spin-3 XXZ Heisenberg model. The build up of quantum correlation during the dynamics, specially as the angle approaches $\pi/2$, is supported  by comparison with an improved numerical quantum phase-space method and further confirmed by the observation that our isolated system thermalizes under its own dynamics, reaching a steady state consistent with the one extracted from a thermal ensemble with a temperature dictated from the system's energy. This indicates a scenario of quantum thermalization which is tied to the growth of entanglement entropy. Although direct experimental measurements  of the Renyi entropy in our macroscopic system  are unfeasible, the excellent agreement  with the theory, which can compute this entropy, does indicate  entanglement  build-up..

\end{abstract}

\date{\today}
\maketitle

Ultra-cold atomic systems featuring long-range interactions   are becoming ideal platforms for probing strongly correlated out-of-equilibrium quantum behavior and, in particular, the phenomenon of quantum magnetism, where magnetic moments with quantized energy levels (spins) interact with one another \cite{Bohn2017,Blatt2012,Saffman2010}. Their  appeal stems from the fact that they  feature  internal levels  that can be  initialized in pure states and  coherently evolved with controllable long-range interactions even under frozen conditions. Recently great advances  have been  accomplished, but,  so far  have been mostly limited to small systems (hundreds or fewer particles) \cite{Bohnet2016,Garttner2017,Richerme2014,Jurcevic2014,Jurcevic2017,Hess2017,Zeiher2017,labuhn2016,Bernien2017},  or to dilute disordered molecular ensembles\cite{Yan2013,Hazzard2014b}. Magnetic quantum dipoles featuring  sizable magnetic moments   offer unique opportunities  since magnetic interactions can directly happen  in an  enlarged set of low-lying hyperfine Zeeman levels  and are not forbidden  by parity and time-reversal symmetry as is the case with  electric dipoles\cite{Lahaye2009}. They  offer untapped opportunities as a quantum resource since $S>1/2$  spin models have more complexity and cost exponentially more resources to classically simulate\cite{Aharonov2007,Hallgren2013}. In fact, the exploration of the complex non-equilibrium dynamics of  dipolar-coupled $S>1/2$ spin models remains a fascinating territory which only starts to be explored \cite{depaz2013,depaz2016}. Here we make a step forward  and report on experimental observations on  how the seven Zeeman populations of an initial  spin coherent state made of $S=3$ spin particles, evolve  and  at long times approach a steady state that is  captured by a statistical ensemble with nonzero thermodynamic entropy, as a result of  the entanglement accumulated during the dynamics.

\begin{figure*}
\centering
\includegraphics[width=15 cm]{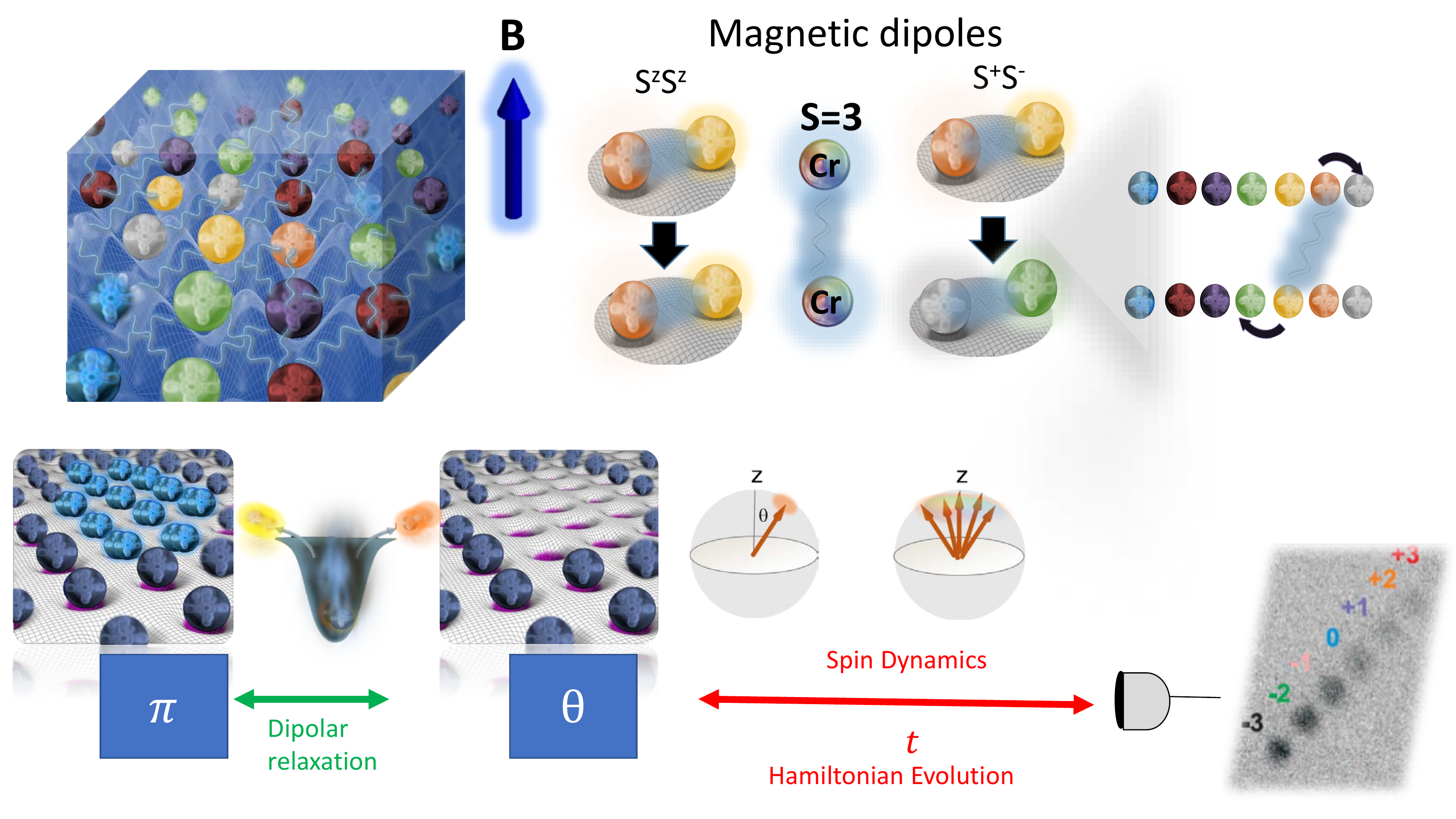}
\caption{{  {\em Sketch of the experiment}. We consider an assembly of $S=3$  Cr atoms in  an optical lattice prepared in  a Mott-insulating state. Dynamics is driven by dipole-dipole interactions which feature both  Ising  ($\hat S_i^z \hat S_j^z$) and exchange  ($\hat S_i^+ \hat S_j^- + h.c$) terms. A first $\pi$ pulse is used to promote all atoms to the most excited spin state. Dipolar relaxation empties doubly occupied sites. Once this is achieved, a second pulse collectively rotates  all spins by an angle $\theta$  from the $B$ field which sets the quantization direction. We then study spin dynamics due to intersite dipole-dipole interactions  by registering the relative population of  the different Zeeman states.}}
\label{principle}
\end{figure*}

In our system the spin degree of freedom is encoded in the Zeeman levels of the purely electronic $S=3$ ground state of $^{52}$Cr atoms. The experiment starts with the production of a spin-3 Bose-Einstein condensate (BEC) of approximately $ 4 \times 10^4$  atoms in the $m_S=-3$ state, following the procedure  described in Ref. \cite{lepoutre2017}. We then adiabatically load the BEC into a three dimensional (3D) optical lattice made by laser beams at 532 nm  \cite{depaz2013}. The lattice structure is rectangular in the horizontal plane, and uses a standard retro-reflecting scheme on the vertical axis (Methods). After loading the atoms into a deep optical lattice, the sample forms  a Mott insulator consisting of a core with doubly-occupied sites ($\overline{n}=2$), surrounded by a 3D shell of singly occupied sites ($\overline{n}=1$), see Fig. \ref{principle}.

The experimental procedure to induce spin dynamics is shown in Fig. \ref{principle}. We initialize the system in a well characterized state consisting of  a macroscopic array of long-lived singly-occupied sites close to unit filling by performing first a filtering protocol. It relies on dipolar relaxation \cite{depaz2013} to empty all doubly-occupied sites within the $\overline{n}=2$ Mott core after the application of  a  $\pi$ rf-pulse that promotes the atoms to the most energetic spin state $m_S=3$. The filtering protocol  takes about 7 ms (Methods). To trigger the spin dynamics we then apply a second rf pulse. This rotates the coherent spin state, such that it forms an angle $\theta$ with respect to the magnetic field which sets the quantization axis (see Fig. \ref{principle}). This prepares a tilted spin coherent state.
The spin dynamics is studied by monitoring the time evolution of the population of the different Zeeman states, using absorption imaging after a Stern Gerlach separation procedure \cite{lepoutre2017}.

A unit filled array of frozen magnetic dipoles in a lattice interact via  dipolar exchange interactions.  In the presence of a  magnetic $B$ field strong  enough to generate Zeeman splittings larger than nearest-neighbor dipolar exchange processes,  only those processes that conserve the total magnetization are energetically allowed  and the dynamics is described by  the following secular Hamiltonian\cite{depaz2013}:
\begin{equation}
\hat H=\sum_{i> j}^N V_{ij} \left[ \hat S_i^z \hat S_j^z -\frac{1}{2} \left( \hat S_i^x \hat S_j^x + \hat S_i^y \hat S_j^y \right) \right]
\label{secular}
\end{equation} where the sum runs over the $N$ populated singly-occupied lattice sites.
It corresponds to a XXZ Heisenberg model with dipolar couplings $V_{i,j}\equiv \frac{\mu_0 (g \mu_B)^2}{4 \pi} \left( \frac{1-3 \cos ^2 \phi_{(i,j)}}{r_{(i,j)}^3}\right)$.  Here $\mu_0$ is the magnetic permeability of vacuum, $g \simeq 2$ is the Land\'e factor, and $\mu_B$ the Bohr magneton. The sum runs over all pairs of particles ($i$,$j$). $r_{(i,j)}$ is the distance between atoms, and $\phi_{(i,j)}$ the angle between their inter-atomic axis and the external magnetic field. The Hamiltonian is given in terms of   spin-3 angular momentum operators, ${\bf{\hat S}}_i=\{\hat S_i^x,\hat S_i^y,\hat S_i^z \}$ , associated to atom $i$.

\begin{figure*}[t]
\centering
\includegraphics[width= 18 cm ]{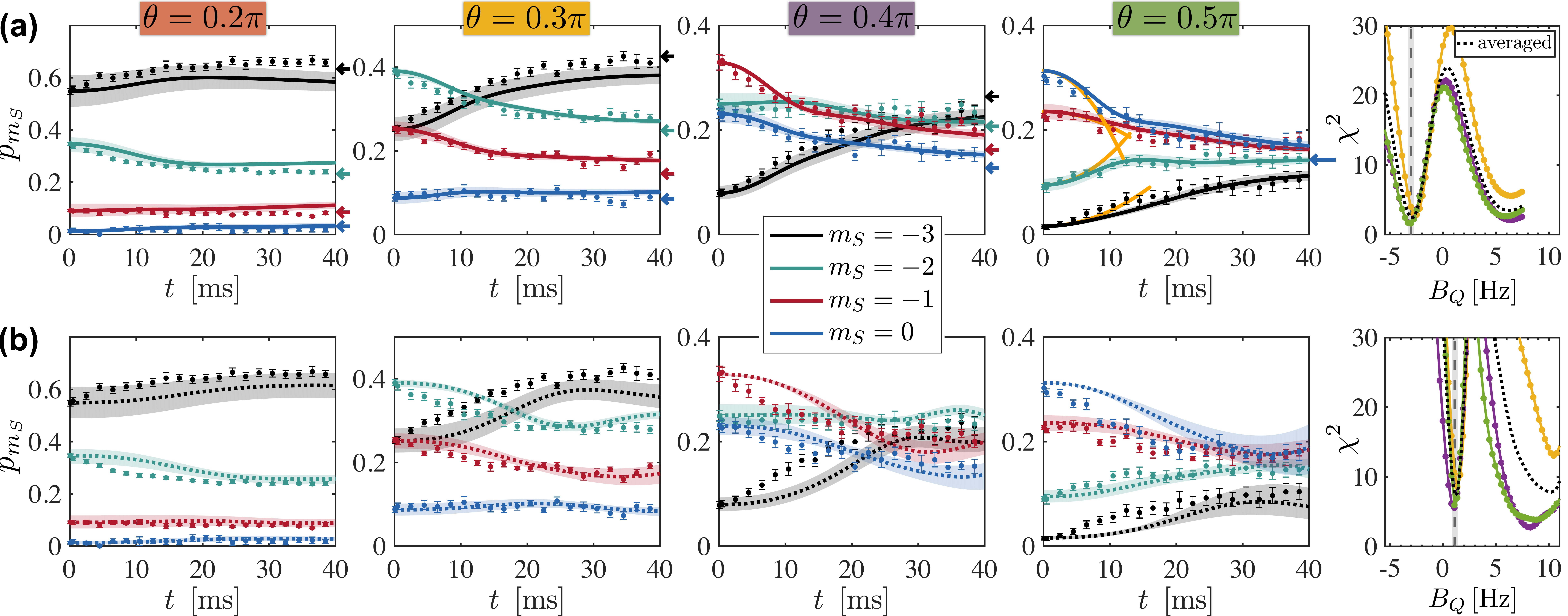}
\caption{{{\em Comparison between classical and quantum dynamics} of the four lowest spin-level populations, $p_{m_S}$, for various initial tilting angles $\theta=0.2\pi$, $0.3\pi$, $0,4\pi$, and $0.5\pi$. (a) Comparison of experimental data with GDTWA simulations (solid lines) on a $7 \times 3 \times 7$ cluster allowing the quadratic Zeeman field $B_Q$ to be the only fitting parameter [here: $B_Q=-3.0\,\text{Hz}$]. (b) Comparison of the experiment with the classical mean-field results (dotted lines) [here: $B_Q=1.1\,\text{Hz}$]. The two plots on the right show quantitative comparison between data and simulations for $\theta=0.3\pi$, $0,4\pi$, and $0.5\pi$ with a reduced $\chi^2$ criteria, for different values of $B_Q$. We excluded the $\theta=0.2\pi$ case here since it shows no significance dynamics.  The best agreement with GDTWA simulations is a factor of three better than with classical simulations. The deviation with the classical simulations is most obvious at short times, and clearly increases with increasing $\theta$. The thick arrows indicate the expected equilibrium population maximizing entropy, for each angle. The red solid line in panel (a) (for $\theta=0.5\pi$) is the result of the perturbative expansion, Eq.(\ref{pert}). The shaded area indicates the range of variation of the populations for evolutions with $\Delta B_Q=\pm 0.3\, \text{Hz}$ and uncertainties in the tilting angles with $\theta = (0.2 \pm 0.018\pi), (0.3 \pm 0.012)\pi, (0.4 \pm 0.012)\pi, (0.5 \pm 0.01)\pi$  (estimated from the experiment).} \label{data_theo}}
\end{figure*}

An important  feature is that the dynamical redistribution of populations can happen for large spins ( $S>1/2$), even though both the  total particle number $N$ and the collective  magnetization $M=   \langle \hat S^{z}\rangle$ are conserved quantities (with  $\hat S^{x,y,z} =\sum_{j=1}^N\hat S_i^{x,y,z}$). The  magnetic dipolar interaction energy  between  $S=3$ spins is 36 times larger than the one for  $S=1/2$ alkali atoms, allowing us to probe such population dynamics at milliseconds time scales, as seen in Fig \ref{data_theo}.

We will first introduce the expected basic dynamical features according to time-dependent perturbation theory. We will focus on the main differences when assuming a classical behavior, or when taking into account quantum correlations. The simplest possible picture  for the population dynamics relies on  a mean field treatment( $i.e.$ neglecting quantum correlations), where  each atom undergoes Larmor precession around an  effective dipolar field created by all the other spins, $\hat{H}^{MF}=\sum_{i=1}^N {\bf B}^{\rm eff}_i  \cdot {\bf \hat {S}}_i$, with  ${\bf B}^{\rm eff}_{i}=-\sum_{j=1}^N \frac{V_{ij}}{2} \{ \langle \hat S_j^x\rangle , \langle \hat S_j^y\rangle,-2\langle \hat S_j^z\rangle\}$. Time-dependent perturbation theory  yields the following equation for $p_{m_S}$, the relative population of Zeeman level $m_S=-3,\dots,3$:
\begin{equation}
p_{m_S}^{\rm MF}(t)=p_{m_S}(0) + \sin[\theta]^4 \alpha_{m_S}(\theta) t^2 {   \mathcal K}_d (t)+ \mathcal{O}\Big(t^6 V_{ij}^6\Big),
\label{pertcl}
\end{equation} We give in Methods exact formulas for  $\alpha_{m_S}(\theta)$. For instance, $\alpha_{m_S=\{-3,-2,-1,0,1,2,3\}}(\pi/2)=135/512 \times \{1,2,-1,-4,-1,2,1\}$. Here ${\mathcal{K }_d}(t)\equiv\frac{t^2}{2N}\sum_{i=1}^N\left[\sum_{j\neq i }^N   V_{ij} B^{dih}_{ij}\right]^2$. Thus classical dynamics is driven by the dipolar field $B^{dih}_{ij}=-9/2\sum_{k\neq j,i}^N(V_{ki}-V_{kj})\cos\theta$.  For a homogeneous gas, $B^{dih}$ vanishes, and the population dynamics with it. This behavior remains valid  at all times given that,  by preparation, all spins point along the same direction initially, they precess around the same classical dipolar field, and thus evolve identically. Therefore, the local magnetization $\langle S_i^z\rangle=M/N$ remains constant for each spin, canceling population dynamics altogether. On the other hand,   in a trapped gas, the inhomogeneous dipolar field introduces a differential precession rate between spins, which results in population dynamics. Note that  $B^{dih}$  is determined  by border effects  and also that ${\mathcal{K }_d}(t)$ itself is time dependent ($ \propto t^2$). Therefore, classically, population redistribution is a slow  $t^4$ process. We emphasize that $B^{dih}$ is proportional to $\cos\theta$  and thus vanishes when $\theta=\pi/2$ where no mean-field dynamics takes place at all.

Quantum fluctuations can drastically modify this behavior and  induce much faster population dynamics even for a homogeneous gas \cite{Hazzard2013}.  Second-order time-dependent perturbation theory on the exact Hamiltonian\cite{NMRbook,Hazzard2014a} (Eq. 1) yields:
\begin{equation}
p_{m_S}(t)=p_{m_S}(0) + \sin[\theta]^4 \alpha_{m_S}(\theta) t^2 V_{\rm eff}^2 + \mathcal{O}\Big( t^4 V_{ij}^4\Big).
\label{pert}
\end{equation} In  contrast to the mean field case, the dynamics grows as $t^2$ and is driven by $V_{\rm eff}\equiv\sqrt{\sum_{i,j\neq i}^N V_{ij}^2/N}$. We emphasize the relative fast decay  of $V_{\rm eff}^2$ with interparticle distance $r$  (as $r^{-6}$), which makes the short time  evolution  mainly  determined by the nearest-neighbor interactions. As $V_{\rm eff}$ is independent of $\theta$  the tipping  angle $\theta$  provides us a way to study an out-of-equilibrium magnetism increasingly determined by quantum correlations as $\theta\rightarrow \pi/2$.

In the experiment, external systematics such as quadratic Zeeman fields, $B_Q$, generated  by tensorial light shifts induced by the lattice lasers --with eigenenergies $B_{Q} m_S^2$-- or  inhomogeneities associated to magnetic field gradients, $\Delta_{ij}=B_i-B_j$ need to be accounted for.  Their role on the short time  dynamics can be understood using perturbation theory (Methods). Quadratic Zeeman fields  can be accounted for by replacing ${\mathcal K}_d (t)\to  {\mathcal K}_d (t)-4/3 Q^2$  and  $ V_{\rm eff}^2 \to V_{\rm eff}^2 -4/3 Q^2$  in the classical and quantum cases respectively, with $Q^2\equiv \frac{1}{N}  \sum_{j\neq i}^N V_{ij} B_Q$. At the mean-field level   $\Delta_{ij}$ directly renormalizes  $B^{dih}_{ij}\to  B^{dih}_{ij}+\Delta_{ij}$. Thus   dipolar inhomogeneities and magnetic field gradients  are  in direct competition. In the quantum case,  magnetic field gradients also enter as  $t^4$  but in this case they play a subdominant role since the leading dipolar dynamics is significantly faster ($\propto t^2$).

Although perturbation theory allows emphasizing some of the main qualitative differences in the classical and in the quantum regime, to accurately describe the population dynamics we need
to go beyond perturbation theory. To accomplish that we parameterize each spin $i$ by a generalized Bloch vector, $\vec \lambda^{[i]}$. In contrast to spin-1/2 systems, this vector  is a 48-dimensional object that  determines all independent elements of the $7 \times 7 (=(2S+1)^2)$ individual spin-3 density matrices, $\hat{\rho}_i(\vec \lambda^{[i]})$, \cite{Bertlmann2008}. Inserting the product state ansatz of the system density matrix, $\hat \rho=\prod_{i=1}^N\hat{\rho}_i(\vec \lambda^{[i]})$, into the von Neumann equation, $d\hat \rho /dt = (-{\rm i}/\hbar) [H,\hat \rho]$ , yields $N\times 48$ independent non-linear mean field equations, in which each generalized Bloch vector evolves in the field of the others (Methods and Supplementary material). The mean field ``classical'' results are obtained by numerically integrating these equations of motion.

To capture the build up of quantum correlations  we developed a generalization of a  semiclassical method  (generalized  discrete truncated Wigner approximation, GDTWA) based on a discrete Monte Carlo sampling in phase space originally derived in the framework of the so-called truncated Wigner approximation (TWA)\cite{Polkovnikov}.  It describes the initial state in terms of a  probability distribution. Initial spin coherent states are ideal since they can be fully described by a positive {\em discrete} probability distribution. For spin-1/2 systems, randomly sampling this initial ``Wigner function'', leads to the discrete truncated Wigner approximation (DTWA) \cite{Schachenmayer2015a,Schachenmayer2015b,Orioli2017}, an approximation that has been remarkably successful and can capture complex  quantum aspects of spin-dynamics. In contrast to the spin-1/2 case,  here (GDTWA) the discrete probabilities are not provided by the eigenvalues of the three Pauli matrices, but instead by the  eigenvalues of the  corresponding 48 generalized SU(7) generators (Methods).

We now describe how our data compares with simulations for different values of $\theta$. In Fig. \ref{data_theo} we show  our data and the comparisons to both the classical and the GDTWA models. The theoretical models take into account the 3D lattice structure and the measured magnetic field gradients along all three directions. We also include the weak quadratic Zeeman field  present in the experiment. Since we cannot measure it directly we allow it to be a fitting parameter;
For each of the four tilting angles used for the measurements  we plot the evolution of fractional populations in different Zeeman states. We only plot the most relevant Zeeman states (the most populated for most of the angles, and only the negative Zeeman states for the symmetric case $\pi/2$). Experimentally, we find that the amplitude of spin dynamics (i.e. the amplitude of the variations of the populations in the different spin states) is stronger when the angle increases.

\begin{figure*}
\centering
\includegraphics[width=10 cm]{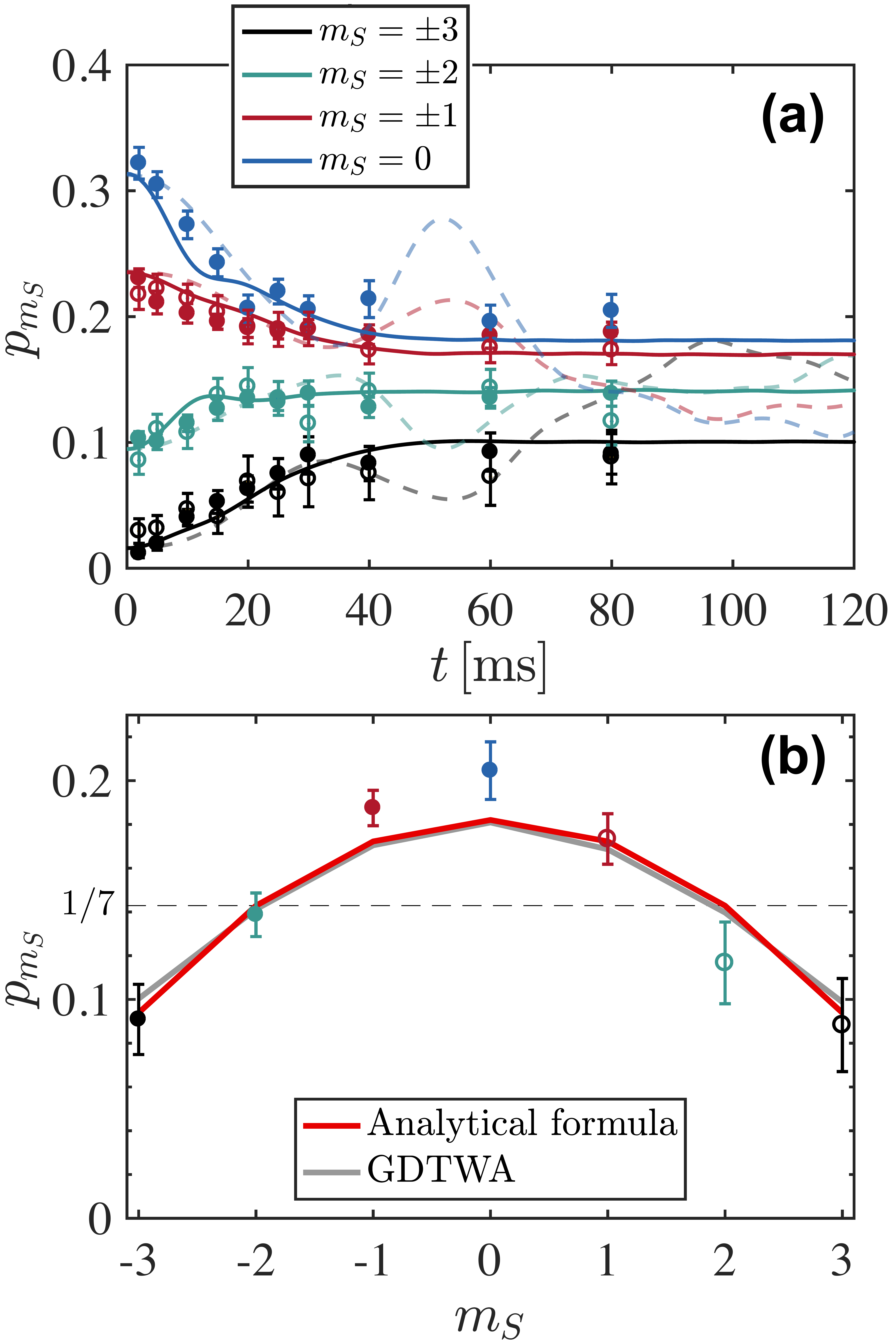}
\caption{{ {\em Quantum Thermalization.} (a) Long-time evolution of state populations for $\theta=\pi/2$: Long time experimental data points are compared to corresponding best fitting GDTWA (solid lines, $B_Q\approx -3.6\,$Hz) and mean-field (thin dashed lines, $B_Q\approx 1.1\,$Hz). (b) The experimental data points at $t = 80\,$ ms are compared to the GDTWA prediction (including field gradients), and the analytical quantum thermalization expression, Eq.\ref{ETHp}, which corresponds to an effective temperature of $-2.5\,$nK, using $B_Q =-3.6\,$Hz.}}
\label{thermal}
\end{figure*}

At small angles the experimental data is qualitatively reproduced by both classical and GDTWA simulations. As can be seen in Fig. \ref{data_theo}, both simulations then yield similar results, but nevertheless show systematic differences. This shows that even at the smallest angles that we have probed, beyond mean-field effects are in principle already at play. However, given the signal-to-noise ratio in the experimental data, it is difficult to quantify the contribution of beyond mean field effects to the spin dynamics at weak rotations. When increasing the angle, it becomes increasingly clear that only the beyond mean-field simulation accounts for the observed dynamics, both at short times ($t < 20$ ms, see Fig \ref{data_theo}), and at long times ($t>40$ ms, see Fig \ref{thermal}a)).

We have performed a systematic study in our numerical simulations, by varying the size of the system which we use. We find that a good agreement between experiment and beyond mean-field theory is only reached provided the number of interacting spins in the simulation is larger than about 60. Taken together, these data show that spin dynamics after the initial quench is inherently many-body, and beyond the grasp of mean-field models. As can also be seen in Fig. \ref{data_theo}, our experimental data at short times is also in excellent agreement with the exact dynamics calculated within the framework of second-order time-dependent perturbation theory (see Eq. (\ref{pert})). Also in good agreement with this equation, we find the dynamics at short times to be roughly independent of the magnetic field gradient applied to the sample (up to values $>30$ MHz$/$m). In contrast, we point out that the experimental data at short times systematically shows faster dynamics than predicted in the classical picture, whose initial $t^4$ dependence (See Eq. 2) fails reproducing the experimental observations.

For an isolated system, entanglement build-up after a quench into a non-equilibrium situation is tied to the scenario of quantum thermalization.  To support the relevance of quantum correlations during dynamics, we thus analyze the long time behavior of the populations. For all tipping angles $\theta$, we observe that the experimental system approaches a steady state,  which is in agreement with predictions of closed system quantum thermalization, given e.g.~by the Eigenstate Thermalization Hypothesis (ETH) \cite{Alessio2016, Kaufman2016}. In particular, we find that the long-time average populations are very well described by the effective thermal distribution  $\hat \rho_{cT}(\beta,\mu) =\frac{e^{-\beta {\hat H}_T -\mu {\hat S}^z}}{{\rm tr}[e^{-\beta {\hat H}_T-\mu \hat{\hat S}^z}]}$ where the chemical potential $\mu$ and inverse temperature $\beta=1/k_B T $ are set by the energy and magnetization of the initial pure state:
\begin{equation}
\langle \hat{H_t}\rangle= {\rm tr}[\hat \rho_{cT} (\beta,\mu)\hat{H_T}] \quad \quad  \langle \hat{S}^z\rangle= {\rm tr}[\hat \rho_{cT} (\beta,\mu) \hat{S}^z],
\label{ETH}
\end{equation} which are conserved throughout the evolution. Here $\hat H_T= \hat H + \sum_i B_Q (\hat S_i^z)^2$ is the total Hamiltonian. As shown  in Fig \ref{data_theo} the steady state populations approach the ones (indicated by the arrows for all tipping angles) dictated by the thermal ensemble when simply setting $\beta=0$ (in which case the maximum entropy state only depends on magnetization).  For angles close to $\pi/2$, however, where quantum effects are most significant, we find a deviation compared to this simplistic prediction. We therefore proceed to  study  this  interesting regime.

\begin{figure*}[t]
\centering
\includegraphics[width= 15 cm]{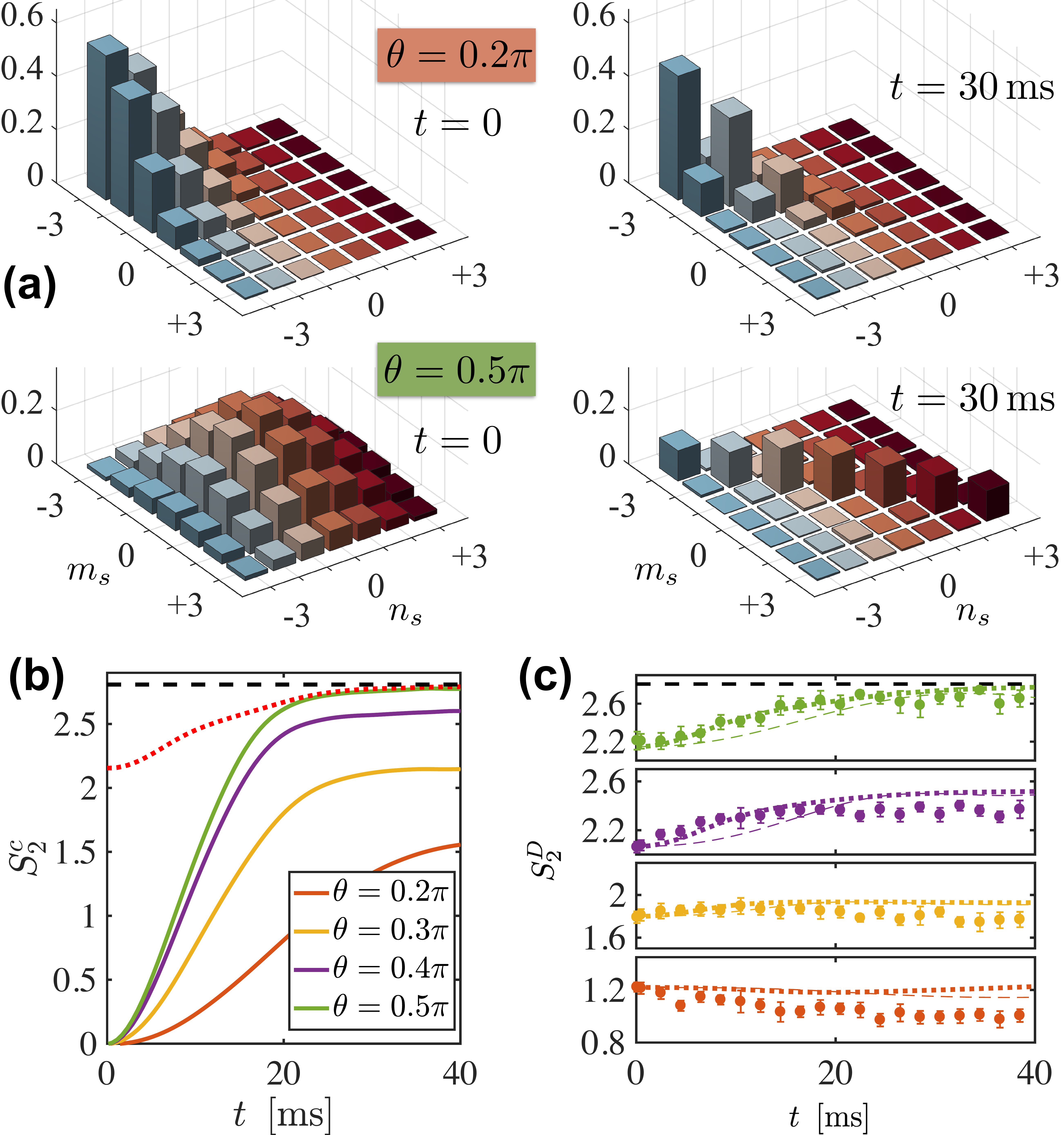}
\caption{ {{\em Entanglement build-up} (a) Absolute values of the central spin density-matrix elements  $|\rho^c_{m_S,n_S}|$ with $\rho^c_{m_s,n_S} \equiv \langle m_S|\hat \rho_{0}|n_S\rangle$,  extracted from  GDTWA simulations with the same parameters as in Fig.~\ref{data_theo}. Off-diagonal single-site coherences are destroyed as the spins become entangled during the quantum dynamics (left two panels: $t=0$, after the tilt, right two panels: after $t=30\,\text{ms}$ evolution). For small  rotation angles  (upper panels: $\theta=0.2\pi$), the system evolves locally into a partially mixed state (un-even spin-state population). For  larger rotation angles (lower panels:  $\theta=0.5\pi$), the local state resembles a maximally  mixed state of the form $\hat \rho_0 \propto\frac{1}{7} \sum_{m_S=-3}^{3} |m_S\rangle\langle m_S\rangle$. (b) Evolution of $S^{c}_2$, the value of the second order Renyi-entropy $S^{(2)}_0$ for the central spin  density matrix. For larger rotation  angles, the  entanglement entropy increases with time, almost reaching the maximum value ($S^{(2) {\rm max}}_0\leq \log_2(7)$, black dashed line). The red dotted line shows the upper bound $S_2^D$, computed only from the diagonal elements (populations) of the average single site density matrix  $\hat \rho_{S}$ (see text) for $\theta = 0.5\pi$. (c) Comparison of the theoretically computed diagonal  entropy (GDTWA: thick dotted line; mean-field: thin dashed line) with the one  reconstructed from the  measured populations.}}
\label{entropy}
\end{figure*}

In Fig. \ref{thermal} we show dynamics up to  longer times,  and confirm the steady state is indeed reached after 40 ms for the $\pi/2$ case, a feature that the GDTWA simulation reproduces, while classical simulations predict an oscillatory behavior for a much longer duration. This qualitative difference between the classical and the quantum behavior is associated with the different origin of thermalization in both pictures: while quantum-mechanically thermalization is tied to the growth of entanglement, classically, reaching a steady state in a system of frozen particles is a consequence of the single particle dephasing in precession induced by field inhomogeneities (which differs from the typical thermalization scenario in mobile particles where  collisions can classically redistribute energies and momenta \cite{Alessio2016,Weiss,Langen207,lev2018}). Our observations clearly rule out a simple mean-field (classical) behavior. Most interestingly, we find a very good agreement between the experimental data points taken at long times, i.e. after the system has reached its steady state, if instead of setting $\beta=0$ we account for the  corrections generated by the quadratic Zeeman field $B_Q$, and  the finite but small energy of the initial state  in Eq. (\ref{ETH}). Using a simple perturbative approach (see Methods and Supplementary material)  we obtain that the $\beta^{(0)}=\mu^{(0)}= 0 $ solutions  should be replaced by:

\begin{eqnarray}
\beta^{(1)}&=& \frac{{\rm tr}[\hat \rho_{cT}  (0,0)\hat H_T]- \langle \hat{H_T}\rangle}{{\rm tr}[\hat \rho_{cT} (0,0) \hat H_T^2]-{\rm tr}[\hat \rho_{cT}  (0,0) \hat H_T]^2 }\nonumber \\
&=&\frac{5 B_Q+9\bar{V}}{24 V_{\rm eff}^2+ 24 B_Q^2},\nonumber \\
p_{m_S}(t_{SS})&\approx& {\rm tr}[\hat \rho_{cT}  (\beta^{(1)},0){\hat p}_{m_S}]=\frac{1}{7}\big(1-\beta^{(1)}B_Q(m_S^2-4)\big),\nonumber \\
\mu^{(1)}=0
\label{ETHp}
\end{eqnarray}
with   $\bar{V} \equiv 1/N\sum_{i> j}^N V_{ij}$. We find $\bar{V}/h\approx -0.57$ Hz and $V_{\rm eff}/h\approx 6.13 $ Hz for our lattice geometry. As $\bar{V}<0$, negative temperatures are expected for low enough $B_Q$ (as allowed for a system whose maximum energy is bounded). Fig. \ref{thermal}b) shows a very good agreement between the equilibrium data and the analytical model. For this comparison, there is no free parameter, since we use the value of $B_Q$ for which the dynamical evolution of the spins is best reproduced by GDTWA simulations. This good agreement confirms the scenario that the coherent Hamiltonian evolution of the many-body system drives it towards a strongly entangled pure state for which the observables display thermal-like behavior. The agreement between the analytical model and the GDTWA at long times shown in Fig. \ref{thermal}b) also indicates that the GDTWA not only captures the short term dynamics (as previously known from the theoretical point of view), but also the approach to equilibrium.

To  compare the analytical formula to the data we have  ignored  magnetic field gradients  in Eq. (\ref{ETHp}). In principle, magnetic field gradients should lead to an equilibrium state where a spatial texture of magnetization develops. However, such a texture requires long-range interactions between remote parts of the cloud, which only occurs for an extremely long time-scale for  dipolar interactions. We have verified (see Supplementary material)  that indeed magnetic field gradients can be neglected to evaluate the quasi-steady state populations reached at 100 ms. This shows that a local equilibrium is first reached, well before the full many-body system may reach true equilibrium with maximum entropy, where all populations in the different Zeeman states would be equally populated.

To further quantify the importance of quantum correlations in the spin dynamics as a function of the tilting angle $\theta$, we analyze from the theoretical point of view the properties of the reduced density matrix for each spin, $\hat \rho_i(\vec \lambda^{[i]})$. In our simulations those density matrices are readily available from the generalized Bloch vectors. To minimize finite size and boundary effects  we focus on the density  matrix of the central spin of our simulated block $\hat \rho_0 $. Even when, as in our simulations, the quantum state of the full system $\hat \rho$ is pure,   the reduced single spin density matrices can assume a mixed character due to the build up of entanglement between the spins. This mixed character is quantified by a reduced purity, ${\rm tr}(\hat \rho_0^2) < 1$ and thus an increased entropy, which we compute in terms of the second-order Renyi entropy, $S_0^{(2)} = -\log_2[{\rm tr}(\hat \rho_0^2)]$. If the state of the full system is pure, ${\rm tr}(\hat \rho^2 )= 1$, the Renyi entropy is a measure of  entanglement \cite{Renyi}: it is zero for product states, and reaches the maximum value of $S_0^{(2){\rm max}}=\log_2[7]$ (the value for a fully mixed state of a spin-3 particle) for many-body states where the quantum information encoded in an individual spin is completely scrambled due to entanglement with  other ones.

As can be seen in Fig. \ref{entropy}, the quantum evolution leads to a growth of $S_0^{(2)}$ already for the smallest investigated angle $\theta = 0.2\pi$. The dynamical growth of entanglement increases significantly for larger tilt angles. At $\theta = \pi/2$, we find that $S_0^{(2)}$ approaches its maximum possible value  $S_0^{(2)\rm {max}}$. Although we cannot perform a full state-tomography from the experimental data, we can compare our experimental data to a ``diagonal entropy'' computed in terms of the diagonal part of the averaged single particle density matrix $\hat \rho_S = (1/N) \sum_{i=1}^N \hat \rho_i(\vec \lambda^{[i]})$. Note that for an homogeneous system $\hat \rho_S =\hat \rho_0$. By neglecting the off-diagonal coherences of $\hat \rho_S$ we can define this entropy as $S_2^D = -\log_2\{{\rm tr}[{\rm diag} (\hat \rho_S)^2]\}$, which  can be readily accessed from the population data, assuming homogeneity: $S_2^D = -\log_2\{ \sum p_{m_S}^2 \}$.

This diagonal entropy is not an entanglement witness, but it provides an upper bound of the entanglement entropy, $S_2^D \geq S_0^{(2)}$. In a translationally
invariant system it increases as  quantum correlations build up with
time and  approaches  the full entropy as the single-spin density
matrices decohere due to entanglement.  However, in our finite system,
boundary effects can obscure this behavior. Although this is the case
at small angles, where diagonal entropy shows a slight reduction as a
function of time, as can be seen in Fig. \ref{entropy} for $\theta\geq 0.3\pi$   we do observe it  increases with
time as the system thermalizes. The non-trivial  growth of the  experimental diagonal entropy  is in  excellent agreement with our theoretical estimates, provided quantum fluctuations are taken into account. Moreover it also  approaches $S_0^{(2)}$ for $\theta=\pi/2$.

In summary our study demonstrates the dominant role of  quantum correlations in the  out-of-equilibrium dynamics of an initially uncorrelated spin coherent state,  when the angle it makes with the magnetic field is close to $\pi/2$.
Our system is the first example of a long-range interacting   many-particle isolated  spin system which internally thermalizes through entanglement build-up, and develops an effective thermal-like behavior through a mechanism which is purely quantum and conservative. The comparison between experiment and theory shows that the GDTWA simulations can be trusted  for studying the dynamics in a complex quantum many-body system, provided a sufficient number of atoms is included in the simulation. Thus, our experiment provides a test-bed for a new theoretical method based on the GDTWA, for the first time applied to systems of large spins, and in a many-body regime where  simulations based on exact diagonalization techniques are intractable with current computational resources. In turn, our study can be used as a first benchmark of  a  quantum simulator of the spin-3 XXZ  Heisenberg model and opens a path towards the study of open problems in quantum many-body physics. For example, by operating the experiment at smaller lattice depths, where tunneling is allowed, we will have the exciting opportunity to study itinerant magnetism, whose description is typically unaccessible to theory, but which is believed to be at the heart of the physics behind high temperature superconductivity \cite{refSupraHT}.

\vspace{2cm}

\noindent\textbf{Acknowledgements}\\
We thank Arghavan Safavi-Naini and Colin Kennedy for their careful reading of the manuscript and useful feedback and Paulo Souto Ribeiro for stimulating discussions about thermalization. The Villetaneuse group acknowledges financial support from Conseil R\'egional d'Ile-de-France under DIM Nano-K / IFRAF, CNRS, Ministère de l'Enseignement Sup\'erieur et de la Recherche within CPER Contract, Universit\'e Sorbonne Paris Cit\'e (USPC), and the Indo-French Centre for the Promotion of Advanced Research - CEFIPRA under the LORIC5404-1 contract. A.M. R acknowledges  supported by NIST, DARPA (W911NF-16-1-0576 through ARO), JILA Physics Frontier Center (NSF-PFC-1125844), AFOSR-MURI, and AFOSR. Work in Strasbourg is supported by IdEx Unistra (project STEMQuS) with funding managed by the French National Research Agency as part of the Investments for the future program. B.Z. acknowledges support of the NSF through a grant to ITAMP.

\newpage

\section{Supplemental Material}

\subsection{Description of the 3D lattice}
The 3D lattice is made with 5 laser beams at 532 nm. On the horizontal plane, 3 beams with the same frequency define a rectangular pattern, with respective directions $\mathbf{u_{H_1}}=\cos(\alpha) \mathbf{u_x}+\sin(\alpha)\mathbf{u_y}$, $\mathbf{u_{H_2}}=-\cos(\alpha) \mathbf{u_x}-\sin(\alpha)\mathbf{u_y}$, $\mathbf{u_{H_3}}=\cos(\pi /4) \mathbf{u_x}+\sin(\pi /4)\mathbf{u_y}$, $\alpha=7/180\pi$. Two other beams, contra-propagating, with a frequency offset by 30 MHz compared to the beams in the horizontal plane, with directions $\mathbf{u_{V_1}}=\cos(\alpha) \mathbf{u_z}+\sin(\alpha)\mathbf{u_x}=-\mathbf{u_{V_2}}$, form an independent light pattern. Calibration of the lattice is performed by standard matter wave diffraction pattern analysis after pulsing lattice beams onto the BEC, with the three pairs of beams $\left(\mathbf {H_1},\mathbf {H_2}\right)$, $\left(\mathbf {H_1},\mathbf {H_3}\right)$ and $\left(\mathbf {V_1},\mathbf {V_2}\right)$. The laser powers are chosen so that these three couples of beams induce almost equal lattice depths, larger than 25 recoil energy. For these lattice depths, the tunneling time is typically 100 ms, and tunneling events can safely be neglected during dynamics.

\subsection{Preparation of a lattice with only singly-occupied sites}

To prepare a lattice of atoms at unit filling, we first slowly load the BEC into a 3D optical lattice, to reach a Mott-insulating state. For our experimental parameters, there exists a core with only doubly-occupied sites, surrounded by a 3D shell of atoms at unit filling. We empty the doubly-occupied sites by performing a rf pulse to promote all atoms from the lowest energy Zeeman state $m_s=-3$ into the state $m_s=3$, which triggers dipolar relaxation. We perform our experiment in presence of a magnetic field which is large enough that dipolar relaxation can be considered as a short-range process \cite{pasquiou2010}. Thus, only atoms in doubly-occupied sites undergo dipolar relaxation, and each dipolar relaxation event empties one doubly-occupied lattice site. We estimate the probability of secondary collisions during this filtering procedure to be below 0.05. After 7 ms, all doubly occupied sites are empty, with about 10 000 remaining atoms.

The spin dynamics experiment is then performed using the atoms remaining in the shell with unit occupancy. Because the sample during dynamics consists of a 3D shell of atoms with unit occupancy within the lattice, border effects might not be fully negligible during dynamics.  Indeed, our estimates is that about 20 percent of the atoms within the shell of singly occupied sites are close to the boundary. It is likely that spin dynamics is slower for these atoms lying close to the frontier of the shell.

Note that the experiment cannot be performed at arbitrarily high magnetic field intensities. As a consequence, some of the atoms which underwent dipolar relaxation remain trapped in very highly excited states of the combined lattice-dipole trap potentials. This translates into losses affecting the sample with unit filling. After 40 ms, from 20 to up to 40 percent of the atoms are typically missing, depending on the magnetic field strength. This phenomenon does not seem to impact the agreement of our spin dynamics data with GDTWA theory as long as losses are below 30 percent.

\subsection{Atom number calibration}

The number of atoms in different spin states is estimated using standard absorption imaging, after spin separation using an applied magnetic field gradient during the free fall of atoms, following a Stern-Gerlach procedure. The cross section for absorption of resonant light strongly depends on the $m_s$ states, through Clebsch-Gordan coefficients.  Therefore, we calibrate the relative sensitivity of the imaging system for the different spin states by comparing the measured populations just after the rf pulse to the theoretically expected values. This calibration depends on the magnetic field direction during spin dynamics, as eddy currents do not allow to rapidly set the direction of the magnetic field during imaging.

For the specific case of $\theta=\pi/2$, we employ a slightly different method to calibrate the different sensitivities. Indeed, the number of atoms in $m_s=+3$ is then very small just after the rf pulse and the detectivity of this Zeeman state is the lowest, due to unfavorable Clebsch-Gordan coefficients. For this specific data set, we thus enforce that the $m_s=-3$ and $m_s=3$ average atom number after spin dynamics are identical. This choice is motivated by the fact that the Hamiltonian preserves magnetization (as experimentally verified for all other data sets), and by the initially symmetric theoretical populations in the different Zeeman states. For example, for the $\pi/2$ data in Fig \ref{data_theo} of the main article the detectivity correction factors of the different Zeeman states are: $f_{-3} = 0.76$, $f_{-2} = 0.96$, $f_{-1} = 1.18$, $f_0 = 1.57$, $f_1 = 2.93$, $f_2 = 2.68$, $f_3 =5.32$.

\subsection{Short time analysis of population dynamics}

Using time-dependent perturbation theory we analyze  the contribution of the different terms in the Hamiltonian  at short times.

\noindent For our system the initial population is given by   $ p_{m_S}(0)=\binom{6}{m_S+3} \Big(\sin \left(\frac{\theta }{2}\right)\Big)^ { (6+2 m_S)}  \Big(\cos \left(\frac{\theta }{2}\right) \Big)^{ (6-2m_S)}$ and the coefficients $\alpha_{m_S}(\theta)$   given by
$\alpha_{-3}(\theta)=\frac{135}{32}\cos^8\left(\frac{\theta}{2}\right)$,
$\alpha_{-2}(\theta)=\frac{135}{32}\cos^6\left(\frac{\theta}{2}\right)[1-3\cos(\theta)]$,
$\alpha_{-1}(\theta)=\frac{135}{256}\cos^4\left(\frac{\theta}{2}\right)[13-20\cos(\theta)+15\cos(2\theta)]$,
$\alpha_{0}(\theta)=\frac{135}{256}\sin^2(\theta)[3+5\cos(2\theta)]$,
$\alpha_{1}(\theta)=\frac{135}{256}\sin^4\left(\frac{\theta}{2}\right)[13+20\cos(\theta)+15\cos(2\theta)]$,
$\alpha_{2}(\theta)=\frac{135}{32}\sin^6\left(\frac{\theta}{2}\right)[1+3\cos(\theta)]$,
$\alpha_{3}(\theta)=\frac{135}{32}\sin^8\left(\frac{\theta}{2}\right)$.

\subsection{Generalized Bloch vectors and the GDTWA}
A generic density matrix for a discrete system with $D$ states on site $i$ takes the form $\hat \rho_{i}=\sum_{\alpha=1,\beta=1}^{D} c_{\alpha,\beta} \ket{\alpha} \bra{\beta}$. For a spin-3 atom $D=7$, and to the states $\ket{\alpha=1,2,3,\dots,6,7}$  we may associate the spin states $\ket{m_S=3,2 \dots,-2,-3}$. Since  $(\hat \rho_{i})^\dag =  \hat \rho_{i}$ and $\text{tr}( \hat \rho_{i})=1$ a total of $D^2-1$ real numbers are needed to describe an arbitrary state. Those numbers can be expressed as expectation values of $D^2-1$ orthogonal observables: $\hat \Lambda^{[i], R}_{\alpha,\beta<\alpha}= \big(\ket{\beta}\bra{\alpha} +  \ket{\alpha}\bra{\beta}\big)$ and $\hat \Lambda^{[i], I}_{\alpha,\beta<\alpha}= -{\rm i}\big(\ket{\beta}\bra{\alpha} - \ket{\alpha}\bra{\beta}\big)$ for $1\leq \alpha \leq D$, $1 \leq \beta <D-1$,  and $\hat \Lambda^{[i],D}_{\alpha}= \sqrt{\frac{2}{{\alpha(\alpha+1)}} }\Big(\sum_{\beta=1}^{\alpha} \ket{\beta}\bra{\beta} - \alpha \ket{\alpha+1}\bra{\alpha+1} \Big)$ for $1\leq \alpha < D-1$. Here, the $\hat \Lambda^{[i],R/I}_{\alpha,\beta<\alpha}$ correspond to measurements of the real ($``R"$) and imaginary ($``I"$) parts of the off-diagonal parts of $c_{\alpha,\beta}$, and $\hat \Lambda^{[i],D}_{\alpha}$ to linear combinations of the real diagonal elements $c_{\alpha,\alpha}$. Together, the set of matrices $\hat \Lambda_\mu^{[i]}\in \{\Lambda^{[i],R/I}_{\alpha,\beta},\hat \Lambda^{[i],D}_{\alpha}\}$  are traceless, $\text{tr}(\hat \Lambda_\mu^{[i]})=0$, and $\text{tr}(\hat \Lambda_\mu^{[i]} \hat \Lambda_\nu^{[i]})=2 \delta_{\mu,\nu}$. Note that for $D=2$, the matrices  reduce to standard Pauli matrices, for $D=3$ to standard Gell-Mann matrices. They are known as generalized Gell-Mann matrices  (GGMs) and are the generators of the SU(D) group \cite{Bertlmann2008}.

The mean-field equations can be written as $(D^2-1) \times N$ coupled non-linear equations for the expectation values of $\lambda^{[i]}_\mu = \langle\hat \Lambda_\mu^{[i]}   \rangle$. The  $\lambda^{[i]}_\mu$ can be interpreted as components of a $D^2-1$ dimensional Bloch vector via the expansion $\hat \rho_{i}(\lambda_\mu^{[i]}) = \big[\mathbb{I}+ \sum_{\mu>0} \lambda_\mu^{[i]} \hat \Lambda_\mu^{[i]}\big]/D$.  We denote the Bloch vector elements associated to the off-diagonal and diagonal GGMs as $\lambda_{\alpha,\beta<\alpha}^{[i], R/I} = (D/2)\,\text{tr}(\hat \Lambda^{[i],R/I}_{\alpha,\beta<\alpha}\hat \rho^{[i]})$ and $\lambda_{\alpha}^{[i], D} = (D/2)\,\text{tr}(\hat \Lambda^{[i],D}_{\alpha}\hat \rho_{i})$, respectively. Furthermore we define $\hat \Lambda^{[i]}_0 ={\mathbb{I}}\sqrt{2/D}$, such that  $\text{tr}(\hat \Lambda_0^{[i]} \hat \Lambda_\nu^{[i]})= 2\delta_{0,\nu}$. Then, an arbitrary operator can be expanded into the orthogonal basis $\{\hat \Lambda_\mu^{[i]}\}$ for $0 \leq \mu < D^2$.  Consider a generic two-spin Hamiltonian between sites $i$, and $j$,  and its expansion into GGMs, $\hat H_{i,j} = \sum_{\mu,\nu} h^{[i,j]}_{\mu,\nu} \hat \Lambda^{[i]}_\mu \hat\Lambda^{[j]}_\nu$. Then the mean-field equations of motion follow from inserting a product-state ansatz $\hat \rho = \prod_i \hat \rho_{i}$ into the von-Neumann equations of motion.  For the Bloch vector at site $i$ ($\hbar = 1$):
$\dot \lambda^{[i]}_\eta$

$\approx  \, \frac{2}{D} \sum_{\mu,\nu,\kappa} h^{[i,j]}_{\mu,\nu} \lambda^{[j]}_\nu  \lambda_\kappa^{[i]}   f_{\mu,\kappa,\eta}$

$\equiv \sum_{\kappa} \mathcal{F}^{[i,j]}_{\eta,\kappa} \lambda_\kappa^{[i]}$.
Here, we defined the ``mean-field matrix''  $\mathcal{F}^{[i,j]}_{\eta,\kappa}\equiv \frac{2}{D} \sum_{\mu,\nu} h^{[i,j]}_{\mu,\nu} \lambda^{[j]}_\nu  f_{\mu,\kappa,\eta}$. Here  the tensor $f_{\mu,\kappa,\eta}$ is defined via $ [ \hat \Lambda^{[i]}_\mu, \hat \Lambda^{[i]}_\kappa] = {\rm i}\, f_{\mu,\kappa,\eta} \hat \Lambda^{[i]}_\eta$, whose elements are the structure constants of the SU(D) group . The full mean-field equations for the generalized Bloch vector at site $i$ are then $\dot \lambda^{[i]}_\eta =  \sum_{\kappa} \Big[ \Big(\sum_j\mathcal{F}^{[i,j]}_{\eta,\kappa}\Big) + h_\kappa^{[i]} \Big] \lambda_\kappa^{[i]}$ where $\hat H^{[i]}=\sum_\kappa h_\kappa^{[i]}  \hat \Lambda^{[i]}_\kappa$ is the expansion of the single-site Hamiltonians containing all local terms (field gradients, quadratic Zeeman fields,etc.) into GGMs. It is straightforward to construct the equations for arbitrary Hamiltonians containing single- and two-site terms numerically, as well as to evolve the generalized Bloch vectors in time.

In the numerical mean-field simulations, the quantum state is represented by $N$ time-dependent generalized Bloch vectors, $\lambda_\mu^{[i]} (t) $.  We evolve the vectors for the initial state $\prod_i \ket{m_S = -3}_i = \prod_i \ket{\alpha = 7}_i $. Explicitly, this state corresponds to a state with $\lambda^{[i],R/I}_{\alpha,\beta<\alpha}(t=0)=0$, $\lambda^{[i],D}_{1,2,3,4,5} (t=0) =0$ and $\lambda^{[i],D}_{6}(t=0)=-\sqrt{21} = -\sqrt{{(D-1)D}/{2}} $. To also simulate dynamics of initially tilted states, i.e.~states created  by applying a unitary collective  rotation, $\ket{\psi_0}=\prod_i \hat U_i(\theta )\ket{m_S = -3}_i$, we simply rotate the equations of motion by rotating the Hamiltonian $\hat H' = \prod_i \hat{ U}_i(\theta)  \hat H  \prod_j  \hat{U}^\dagger_j(\theta)]$.

In contrast, in the GDTWA approach we describe the initial state not by a generalized Bloch vector, but instead by a probability ``Wigner'' distribution, $p^{[i]}_{\mu,a_\mu}$,  for certain discrete configurations of Bloch vector elements, $\lambda_{\mu,a_\mu}^{[i]}$. Initially, the probabilities and configurations are chosen in such a way that on average $\lambda_\mu^{[i]} (t=0)=\sum_{a_\mu} p^{[i]}_{\mu,a_\mu} \lambda_{\mu,a_\mu}^{[i]} \equiv \overline {\lambda_{\mu,a_\mu}^{[i]}} $. In practice, the initial multi-spin configurations are selected via a random sampling of $p^{[i]}_{\mu,a_\mu}$ for each spin $i$ and each Bloch vector component $\mu$. Then the individually selected configurations are evolved according to the non-linear mean-field equations. Observables are computed from a statistical average over the different trajectories. It is important to note that due to the non-linear nature of the equations, this approach can capture the build-up of correlations, e.g.~at later times in general $\overline {\lambda_{\mu}^{[i]}\lambda_{\nu}^{[j]}(t)} \neq \overline {\lambda_{\mu}^{[i]}(t)} \, \overline {\lambda_{\nu}^{[j]}(t)}$.

In particular,  as discrete set of initial configurations, $\{\lambda_{\mu,a_\mu}^{[i]}\}$, we use a set which is inspired by a ``projective measurement, of the GGMs'': For each $\lambda_\mu^{[i]}(t=0)$, we choose a set of initial configurations given by the eigenvalues of each GGM. Consider the eigen-expansion of the GGMs, $\hat\Lambda^{[i]}_{\mu} = \sum_{a_\mu} \eta^{[i]}_{\mu,a_\mu}  \ket{\eta_{\mu,a_\mu}^{[i]}}\bra{\eta_{\mu,a_\mu}^{[i]}}$, where  $\eta_{\mu,a_\mu}^{[i]}$ and $\ket{\eta_{\mu,a_\mu}^{[i]}}$ denote the eigen-values and eigen-vectors, respectively. Then, we choose the ``a-th'' eigenvalue, $\lambda_\mu^{[i]}(t=0) = (D/2)\,\eta_{\mu,a_\mu}^{[i]}$, with probability $p^{[i]}_{\mu,a_\mu} = \text{tr}[\hat \rho_0^{[i]}\ket{\eta_{\mu,a}^{[i]}}\bra{\eta_{\mu,a}^{[i]}}]$, where $\hat \rho^{[i]}_0 = \ket{\alpha = 7}\bra{\alpha = 7}_i$. Note that this choice is  a generalization of the one used for the spin-1/2 DTWA method\cite{Schachenmayer2015a,Schachenmayer2015b}, and for $D=2$, we reproduce the DTWA sampling. Specifically, for the initial state  $\ket{m_S = -3}_i$, this prescription leads to fixed ``diagonal'' Bloch vector elements $\lambda^{[i],D}_{1,2,3,4,5} (t=0) =0$ and $\lambda^{[i],D}_{6}(t=0)=-\sqrt{21}$, fixed off-diagonal elements $\lambda^{[i],R/I}_{\alpha<7,\beta<\alpha} = 0$ and fluctuating off-diagonal elements $\lambda^{[i],R/I}_{\alpha=7,\beta=1,\dots 6} \in \{-D/2,+D/2\}$, each with $50\%$ probability.

\subsection{Quantum thermalization}  It is generally believed that the unitary
quantum evolution of a complex quantum  system leads to an apparent maximum-entropy state
that can be described by thermodynamical ensembles that properly account for the conserved quantities. In our systems those are the energy and magnetization. We thus postulate that the steady state properties of local  observables, such as the relative population of Zeeman levels, can be described in our system  by the thermal distribution $\hat \rho_{cT}(\beta,\mu) =\frac{e^{-\beta {\hat H}_T -\mu {\hat S}^z}}{{\rm tr}[e^{-\beta {\hat H}_T-\mu \hat{\hat S}^z}]}$ where $\mu$ and $\beta=1/k_B T $ are  the chemical potential and inverse temperature set by the energy and magnetization respectively accordingly to Eq.(\ref{ETH}).
While the determination of $\beta$ and $\mu$ can be a challenging task for a complex many-body system, the anisotropic character of the dipolar interactions facilitates an analytic high temperature expansion around $\beta=0$.

Under this assumption, the chemical potential to leading order, is set by  $\langle \hat{S}^z\rangle= {\rm tr}[\hat \rho_{cT} (0,\mu^{(0)}) \hat{S}^z]=\frac{\sum_{m_S=-3}^{3} m_S e^{-\mu^{(0)} m_S}}{\sum_{m_S=-3}^{3} e^{-\mu^{(0)}m_S}}$ and therefore $p_{m_S}^{(0)}(t_{SS})=\frac{e^{-\mu^{(0)} m_S}}{\sum_{m=-3}^{3} e^{-\mu^{(0)}m}}$. Here $t_{SS}$ refers to the steady state.
These are the populations indicated by  arrows in Fig \ref{data_theo}. The case $\theta=\pi/2$ is particularly simple since $\langle \hat{S}_z\rangle=0$  and thus $\mu^{(0)}=0$ and  $p_{m_S}^{(0)}(t_{SS})=1/7$.
This solution however shows deviations with the observed long time dynamics indicating that finite $\beta$ corrections are relevant. To first order in $ \beta ^{(1)} $ the chemical potential can be written as $\mu^{(1)}= \mu^{(0)}+ \beta ^{(1)} \delta \nu $ and  the  solutions of  Eq. (\ref{ETH})  described by the relations:

$\langle \hat{S}^z\rangle= \widetilde{{\hat S}^z}-\beta ^{(1)}\Big (\delta \nu \Delta\widetilde{{\hat S}^z{\hat S}^z}+\Delta\widetilde{{\hat H}_T{\hat S}^z}\Big)$ and $\langle \hat{H}_T\rangle= \widetilde{{\hat H}_T}-\beta ^{(1)}\Big(\delta \nu \Delta\widetilde{{\hat H}_T{\hat S}^z}+\Delta\widetilde{{\hat H}_T{\hat H}_T}\Big)$, where we have defined $\widetilde{\hat {\mathcal O}}\equiv{\rm tr} [\hat \rho_{cT}(0,\mu^{(0)})\hat {\mathcal O}]$ and $\Delta\widetilde{{\hat {\mathcal O}}{\hat { \mathcal A}}}\equiv \widetilde{{\hat {\mathcal O}}{\hat { \mathcal A}}}-\widetilde{\hat {\mathcal O}}\widetilde{\hat { \mathcal A}}$.

Solutions of those equations are particularly simple for the $\theta=\pi/2$ case where $\mu^{(0)}=0$, $\widetilde{{\hat S}^z}=0$, $\Delta\widetilde{{\hat H}_T{\hat S}^z}=0$ and $\Delta\widetilde{{\hat S}^z{\hat S}^z}= N I_2$,
$\widetilde{{\hat H}_T}=N B_Q I_2$ and $\Delta\widetilde{{\hat H}_T\hat{H}_T}=N (B_Q^2 (I_4-I_2^2)+ 12  V_{\rm eff}^2 I_2^2)$ with $I_r=(\sum_{m=-3}^3 m^r )/7$  (thus $I_2=4$ and $I_4=28$). Those yield  the expressions for  $ \beta^{(1)}$ and $\mu^{(1)}$ quoted in Eq.(\ref{ETHp}). In the presence of linear gradients $\hat{H}_T\to \hat{H}_T+\sum_i^N B_{i=1} \hat{S}^z_i$, under the assumption that $\sum_i^N B_{i=1}=0$,  the  inverse temperature equation in Eq.(\ref{ETHp}) for the case of $\theta=\pi/2$ should be replaced by $\beta^{(1)}=\frac{5 B_Q+9\bar{V}}{24 V_{\rm eff}^2+ 24 B_Q^2 + 8 V_{\rm B}^2 }$ with  $  V_{\rm B}^2 =1/N \sum_{i=1}^N B_i^2$.

\begin{figure}\centering
\includegraphics[width= 8.5 cm]{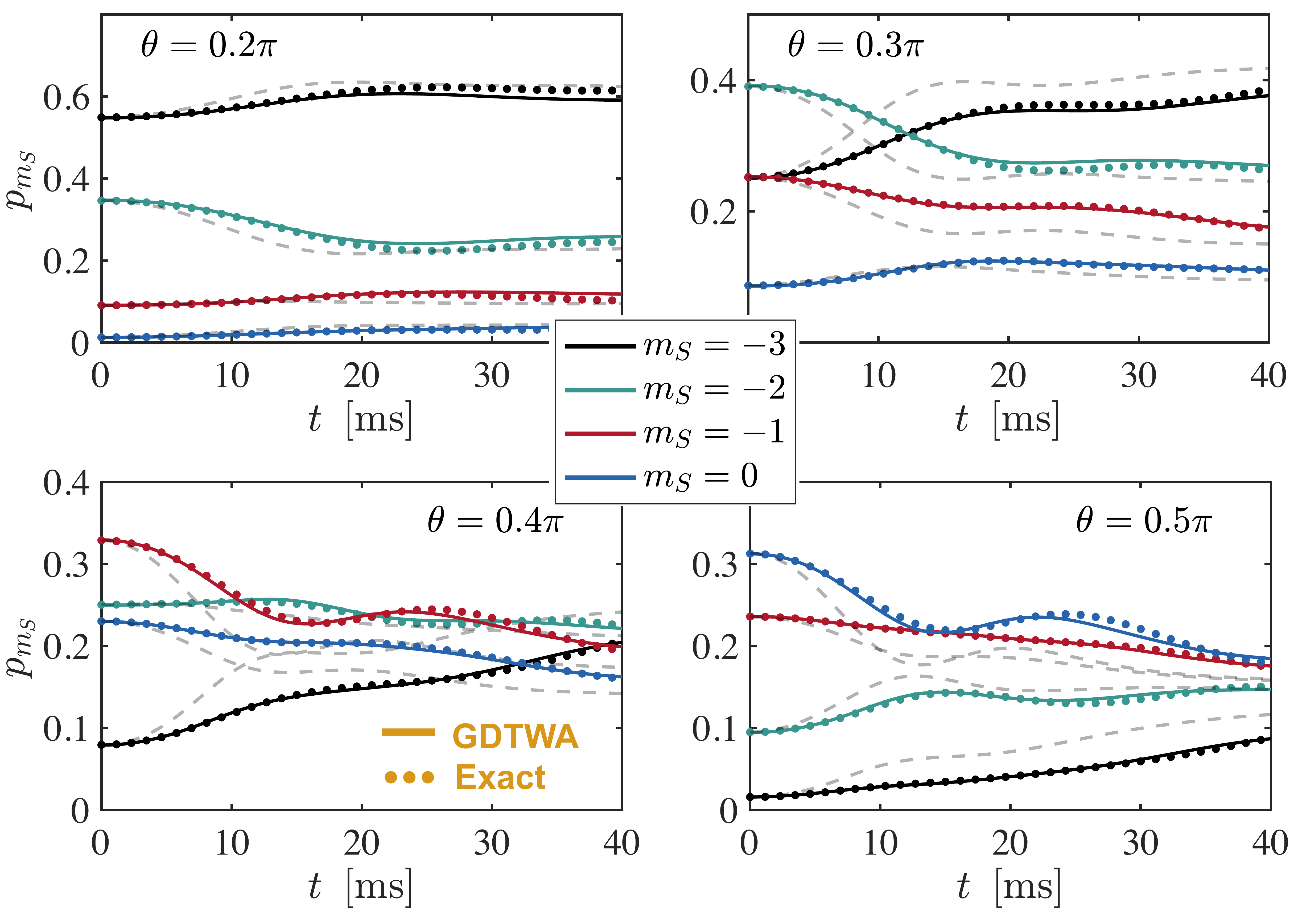}
\caption{Benchmark of the GDTWA on a small plaquette -- Time evolution of state populations for all four different initial tilt angles for the parameters used in the main text on a small  $L_x \times L_y \times L_z = 2 \times 1 \times 4$ plaquette. The exact-diagonalization results (points) are compared to the GDTWA predictions (lines). The GDTWA provides  excellent quantitative predictions on the considered time-scale. As a comparison, the large system GDTWA ($6\times 3\times 6$) results are shown as thin dashed grey lines.}
\label{fig:plaq}
\end{figure}

\subsection{Benchmark of the GDTWA on a small plaquette }

We simulate dynamics of atoms in a 3D lattice geometry. The rapid Hilbert-space growth with system size ($7^N$) prohibits an exact diagonalization simulation in a real 3D system. The lattice constants in the different dimensions are $(d_x, d_y, d_z) \approx  (1.12, 2.24, 1.01) \times \lambda/2$. Due to this geometry and the large spacing along the $y$-direction, dipole-couplings  along the $y$ dimension are relatively small. Thus, to check the validity of our GDTWA simulation, we benchmark it against the exact solution on a small 2D plaquette in the $x-z$ plane. Fig.~\ref{fig:plaq} shows a benchmark of the GDTWA prediction for the population dynamics for all different tilt angles on a $L_x \times L_y \times L_z = 2 \times 1 \times 4$ plaquette (otherwise the same parameters as in the main text are used). The agreement of the GDTWA (lines) with the ED results (points) is remarkable. Furthermore, we note that modeling the experiment with larger 3D system sizes remains crucial, as is seen by the large difference to the GDTWA result for the large $6\times 3\times 6$ system.

\begin{figure}\centering
\includegraphics[width= 8.5 cm]{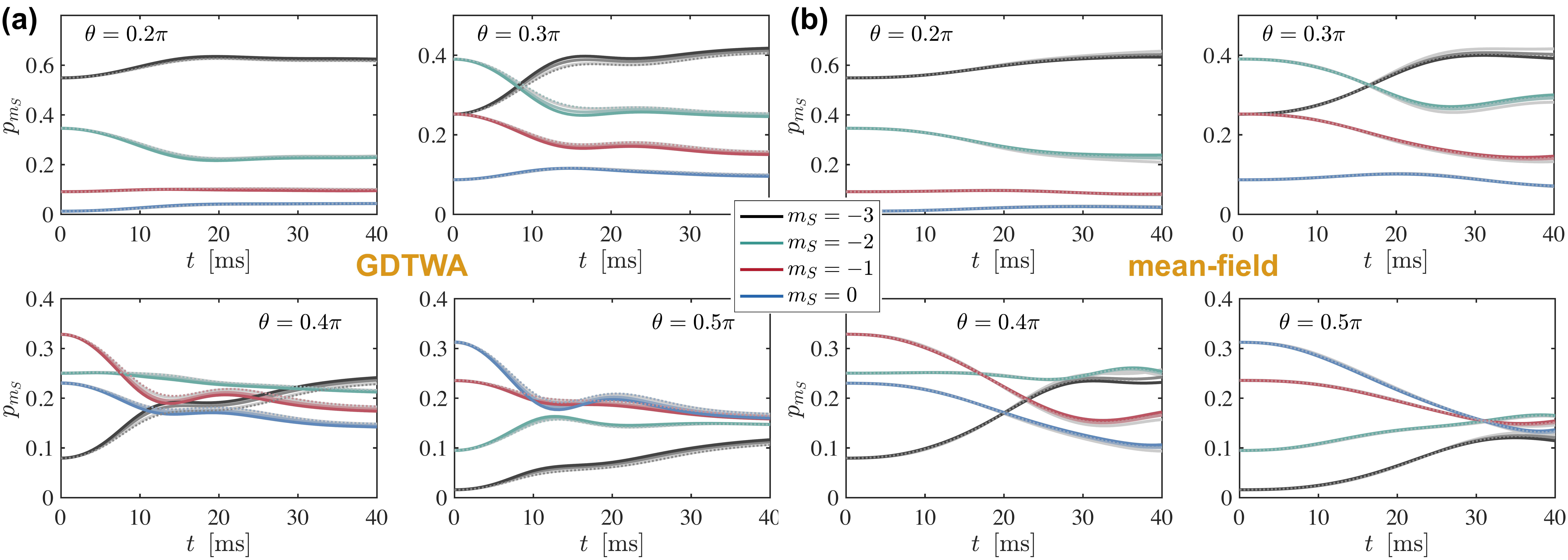}
\caption{System size convergence: Time evolution of state populations for all four different initial tilt angles for the parameters used in the main text. The 3D system size $L_x \times L_y \times L_z$  is increased. (a) Results for GDTWA simualtions ($B_Q=4.4\,\text{Hz}$). Three lines from light to dark color are for $L_y=3$, and $L_x=L_z = 4,5,6$. The dashed lines are for a $4\times4\times4$ system and the overlap of the dashed line with the $4\times 3\times 4$ demonstrates convergence in the $y$ direction. (b) Same results for mean-field simulations ($B_Q=1.4\,\text{Hz}$), light to dark lines are for  $L_y=4$, and $L_x=L_z = 9,11,13$, and dashed lines are for a $11\times 3\times 11$ system.}
\label{fig:sys_conv}
\end{figure}

\subsection{System size convergence in numerical simulations}

Even with the classical equations it is very hard to model the macroscopic number of $10^4$ atoms in the experiment. We therefore simulate the population dynamics of a bulk of atoms by using a small 3D $L_x \times L_y \times L_z$ block of increasing size and checking for finite size convergence. Due to the lattice geometry (with increased spacing in the $y$ direction) finite size convergence is achieved for  values $L_y < L_x,L_z$.  Our finite size comparisons are summarized in  Fig.~\ref{fig:sys_conv} for both our GDTWA and mean-field simulations. In the GDTWA case it is evident that a size of $L_x \times L_y \times L_z = 6\times3\times6$ already give well converged results for all populations and all tilt angles on our time-scale of interest (for the results in the main text we use a $7\times 3 \times 7$ block).

The mean-field result requires much larger system sizes for finite size convergence. This is a direct consequence of the different thermalization mechanism of the mean-field dynamics. Individual spin-density matrices remain pure in the mean-field case throughout the simulation. Finite entropy only builds up in the system-averaged local spin-density matrix, and stems from an inhomogeneous evolution of local phases due to field-gradients and finite-size effects. Therefore large system sizes are required for a converged average. In the main text, we use a system size of $L_x \times L_y \times L_z = 13\times4\times13$, which provides converged results on our time-scale of interest [cf. Fig.~\ref{fig:sys_conv}(b)].

\begin{figure*}\centering
\includegraphics[width= 18 cm]{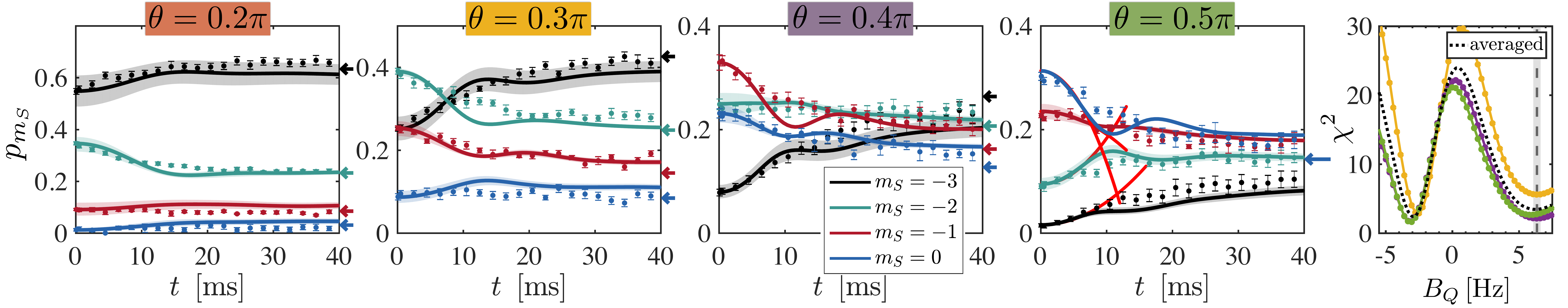}
\caption{Identical to Fig.~2a, but instead of the best fitting $B_Q$ value, the best fitting {\em positive} value of $B_Q$ is chosen: Comparison of experimental data with GDTWA simulations of the four lowest spin-level populations, $p_{m_S}$, for various initial tilting angles $\theta=0.2\pi$, $0.3\pi$, $0,4\pi$, and $0.5\pi$ on a $7 \times 3 \times 7$ cluster allowing the quadratic Zeeman field $B_Q$ to be the only fitting parameter [here: $B_Q=6.3\text{Hz}$]. The red solid line (for $\theta=0.5\pi$) is the result of the perturbative expansion, Eq.~(3). The shaded area indicates the range of variation of the populations for evolutions with $\Delta B_Q=\pm 0.3\, \text{Hz}$ and uncertainties in the tilting angles with $\theta = (0.2 \pm 0.018\pi), (0.3 \pm 0.012)\pi, (0.4 \pm 0.012)\pi, (0.5 \pm 0.01)\pi$  (estimated from the experiment). } \label{compa}
\end{figure*}

\subsection{Details on determination of best quadratic shift }

Here, we provide more details on our determination of the best value of $B_Q$, which we take as only fitting parameter for the full numerical simulations to the experimental data points. We compute the deviation from each experimental data point as $\chi^2_{m_S}(t) =  [p_{m_S}^{[\text{sim.}]}(t) - p_{m_S}^{[\text{exp.}]}(t)]^2/\sigma^2_{m_S}(t) $, with $p_{m_S}^{[\text{sim./exp.}]}(t)$ the simulated and experimental state population, respectively and  $\sigma_{m_S}(t)$ the experimental error bar for the respective data point. For each tilt angle we compute the mean deviation as $\chi^2 = \overline{\chi^2_{m_S}(t) }$, where the average is taken over all data points (in time) and the four spin-populations. This is plotted in Fig \ref{data_theo} of the main text. We excluded  $\theta = 0.2\pi$ since at this low angle there is not significant evolution of the population. We extract the overall best fitting $B_Q$ from the averaged $\chi^2$ over all tilt angles as $B_Q \approx -3 \,\text{Hz}$. In the mean-field case, $\chi^2$ deviates more wildly for large $B_Q$. The relatively best fit for all data points is found for $B_Q \approx 1.1 \,\text{Hz}$. The overall deviation at the best value of $B_Q$ is about three times smaller for the GDTWA than for the mean-field case [$\overline{\chi^2}_{\rm GDTWA} \approx 2.4$ and $\overline{\chi^2}_{\rm MF} \approx 7.3$, respectively]. Just for completeness in Fig.~\ref{compa} we also show the GDTWA dynamics at the best possible positive $B_Q=6.3\,\text{Hz}$, the value at which the  second minimum in $\chi^2$ is obtained (here, $\overline{\chi^2}_{\rm GDTWA} \approx 3.5$). Other parameters are identical to Fig \ref{data_theo} in the main text. Also for this positive value of $B_Q$ the agreement with the experiment is decent, however, some qualitative oscillations from the simulations are not perfectly reproduced by the experiment.

For the different long-time data set used for Fig \ref{thermal} we perform the identical $\chi^2$ optimization using an average taken over all available $7$ Zeeman sub-states but for the shown $\theta = 0.5\pi$ data only. For this different data set we find an optimal $\overline{\chi^2}_{\rm GDTWA} \approx  1.2$ for $B_Q = -3.6\,$Hz, compared to a classical best fit with $\overline{\chi^2}_{\rm MF} \approx 2.9$ for $B_Q = 1.1\,$Hz.

\begin{figure}\centering
\includegraphics[width= 8.5 cm]{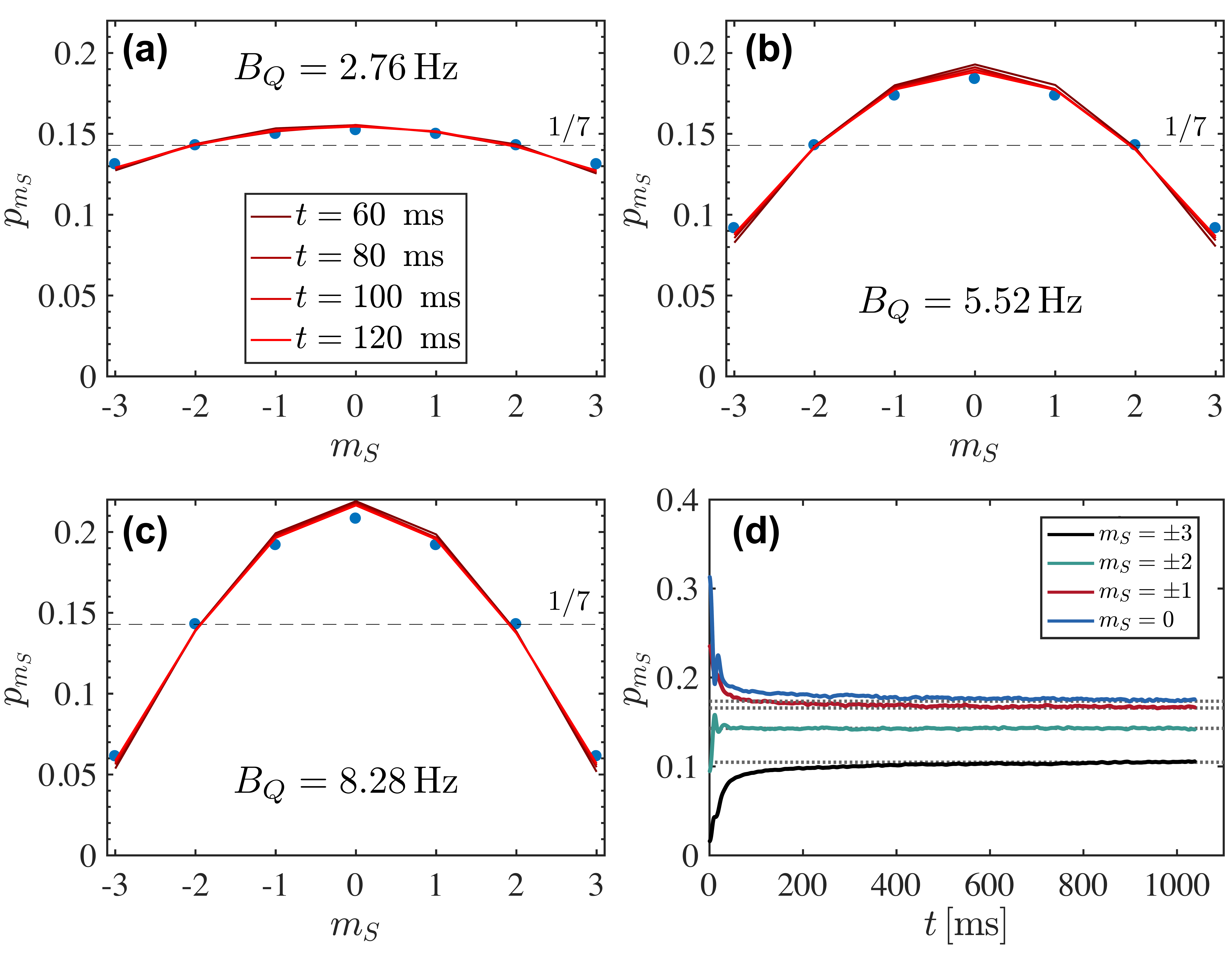}
\caption{GDTWA vs.~analytical formula: (a)-(c) Comparisons of various late time Zeeman level populations for zero magnetic field gradients and different  quadratic tensor light shifts $B_Q=2.76\,$Hz, $B_Q=5.52\,$Hz, and $B_Q=8.28\,$Hz, respectively ($7\times 3\times 7$ lattice). The blue points show the analytical thermalization values from Eq.~(6).  GDTWA results converge to the thermal prediction for all the $B_Q$ cases displayed. (d) Long-time evolution up to $1\,$s with half the field-gradients from the experiment (and $B_Q = 5.52\,$Hz). Horizontal dotted lines show the prediction from Eq.~(6) after including the exact gradients on the $7\times 3\times 7$ block. At the experimentally relevant time-scales ($\sim 80$ms), the population redistribution stems mostly from the interactions and the quadratic tensor light shifts. Only on a much longer scale ($\sim 0.5$s) the redistribution from the field gradients causes a relaxation of the GDTWA simulation to the analytical estimation.}
\label{fig:long}
\end{figure}

\begin{figure}\centering
\includegraphics[width= 8.5 cm]{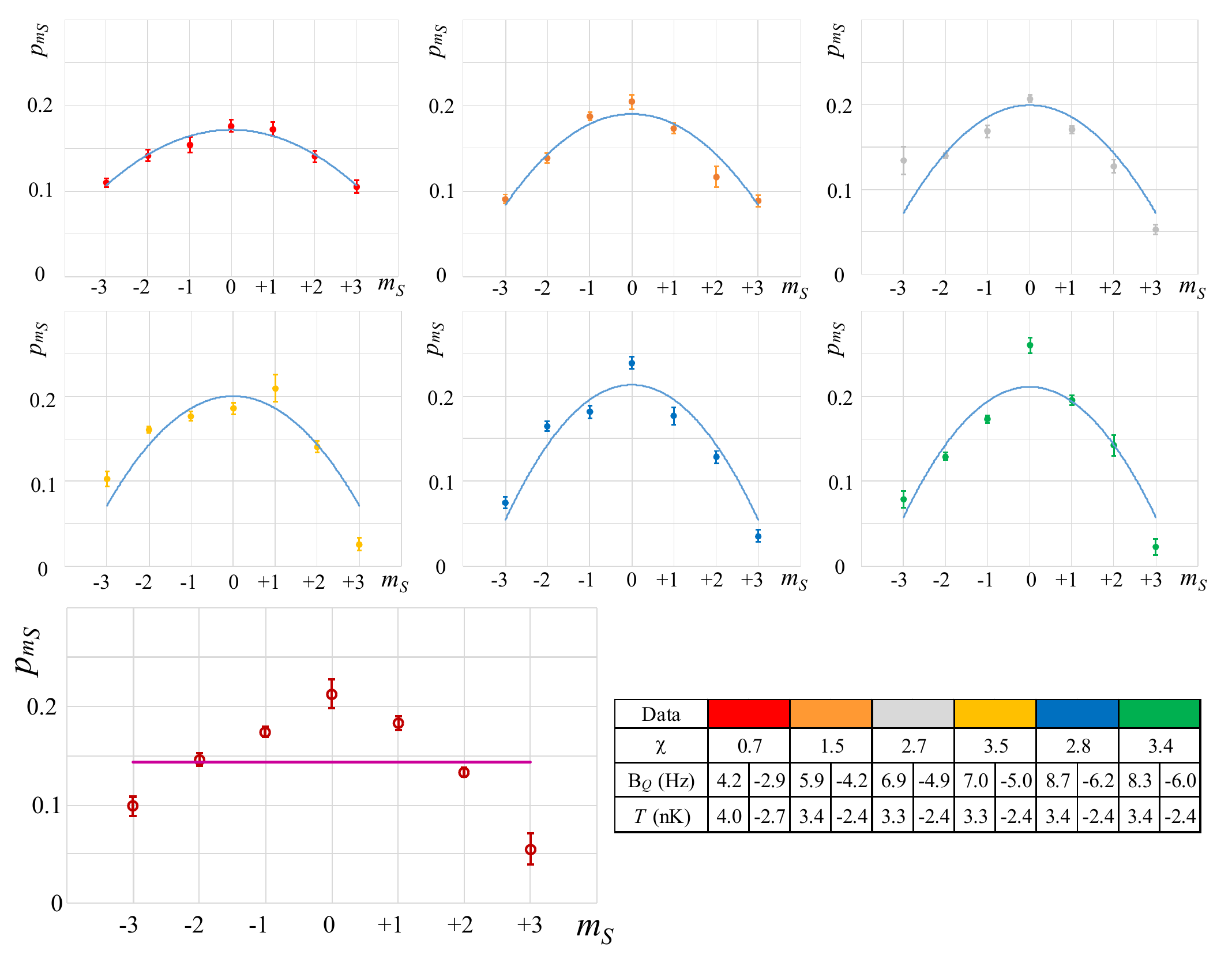}
\caption{Thermalization at long time: extensive data, and comparison with our analytical model. Top 6 figures (filled circles): long time spin populations for 6 different data sets. The full lines correspond to the result of the analytical model with using $B_Q$ as a fitting parameter whose value is displayed in the Table. Bottom left: the empty circles correspond to average of these 6 data, the error bar corresponding to the associated standard deviation. The full line indicates 1/7. Bottom right: for each data the value of $B_Q$ and of the associated  temperature $T$, deduced by comparing to our analytical model, are given, together with the value of  the $\chi^2$ estimator (see above) characterizing the agreement between theory and experiment calculated over all seven spin populations.}
\label{LongTime}
\end{figure}

\subsection{Long-time dynamics and quantum thermalization}

In the main article (see Figure \ref{thermal}) we discuss about the thermalization process occurring in our spin system. We derive an analytic formula which links  the long time populations of the different Zeeman levels  to the ones  of a thermal ensemble with the same energy and magnetization of the initial state. The formula, however, is not exact. It is based on  a high temperature perturbative expansion which is strictly valid in the regime $B_Q\ll \sqrt{V_{\rm eff}}$.  In this section we benchmark the validity of the  formula by comparisons with GDTWA dynamics at different quadratic fields. The summary is presented  in Fig. \ref{fig:long}(a)-(c), where we show that the analytic formula is valid for the experimental range of $B_Q$ parameters.

In addition to the quadratic Zeeman field,  magnetic field gradients  are also present in the experiment. While those terms can affect the thermalization dynamics leading to the development of spatial magnetization textures at equilibrium, we argue in the main text that such a texture is expected to occur at extremely long times in a dipolar interacting system  since it requires  interactions between remote parts of the cloud. Here we  validate such claim  using  GDTWA simulations. As seen in Fig. \ref{fig:long}(d) for the times scales currently accessible in experiments the gradients just marginally flatten out the population distribution compared to the zero gradient case.

With this validation of the analytic formula, we can use it to further characterize the  asymptotic state obtained under different experimental measurements. In Fig. \ref{LongTime}, we show all available long time data for the spin populations. As the quadratic light shift $B_Q$ is very sensitive to the relative intensity of the five different lattice laser beams, we expect variations of  $B_Q$ between  different data sets and therefore, accordingly to the analytic formula, a correspondent variation on the fractional populations $p_{m_S}$ at  long times except  from $p_{\pm 2}$  which should remain  close to $1/7$. All these features are indeed observed  in Fig. \ref{LongTime}.

We use the model to find the $B_Q$ that best fits the data, and deduce from it the temperature $T$ at which thermalization takes place (see Table in Fig. \ref{LongTime}). We point out that, similar to the case shown in Fig \ref{data_theo}, comparing our data to the analytical model typically leads to two possible values of $B_Q$ (one positive, and one negative), for which the $\chi^2$ shows a local minimum. Distinguishing between these two local minima is difficult (i.e. they share a similar $\chi^2$) – which is a consequence that $|B_Q|>  |\bar{V}|$.

\begin{figure}
\centering
\includegraphics[width= 8.5 cm]{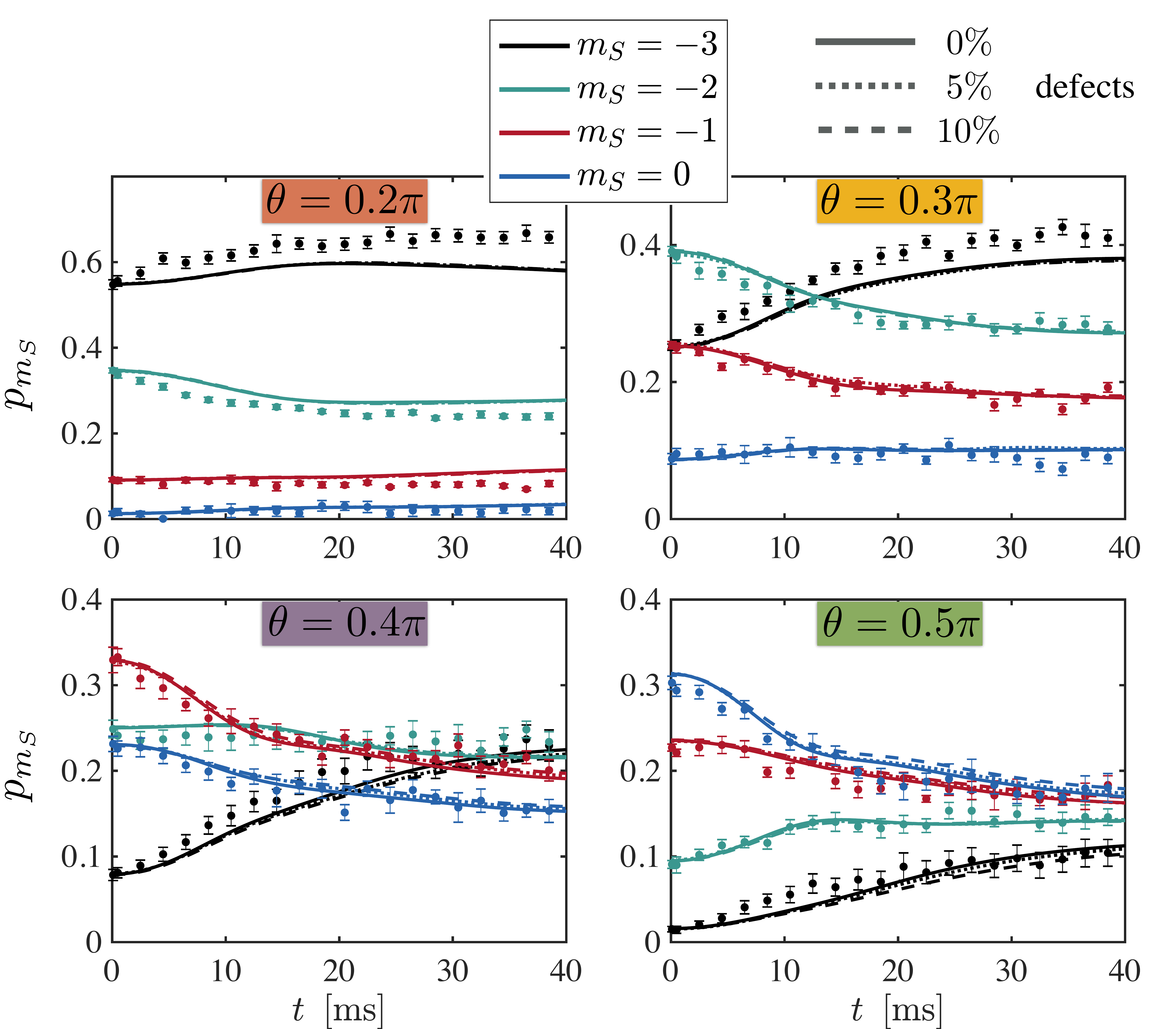}
\caption{{\em The effect of defects} is very small. GDTWA simulations for the populations with identical parameters as in Fig.~2 of the main manuscript. Additionally, dotted and dashed lines show results for $5\%$ and $10\%$ defect density, respectively. The statistical average over $192$ different defect realizations has been taken into account. Differences are more pronounced for larger tilting angles, but still minor.}
\label{holes}
\end{figure}

\begin{figure}
\centering
\includegraphics[width= 8.5 cm]{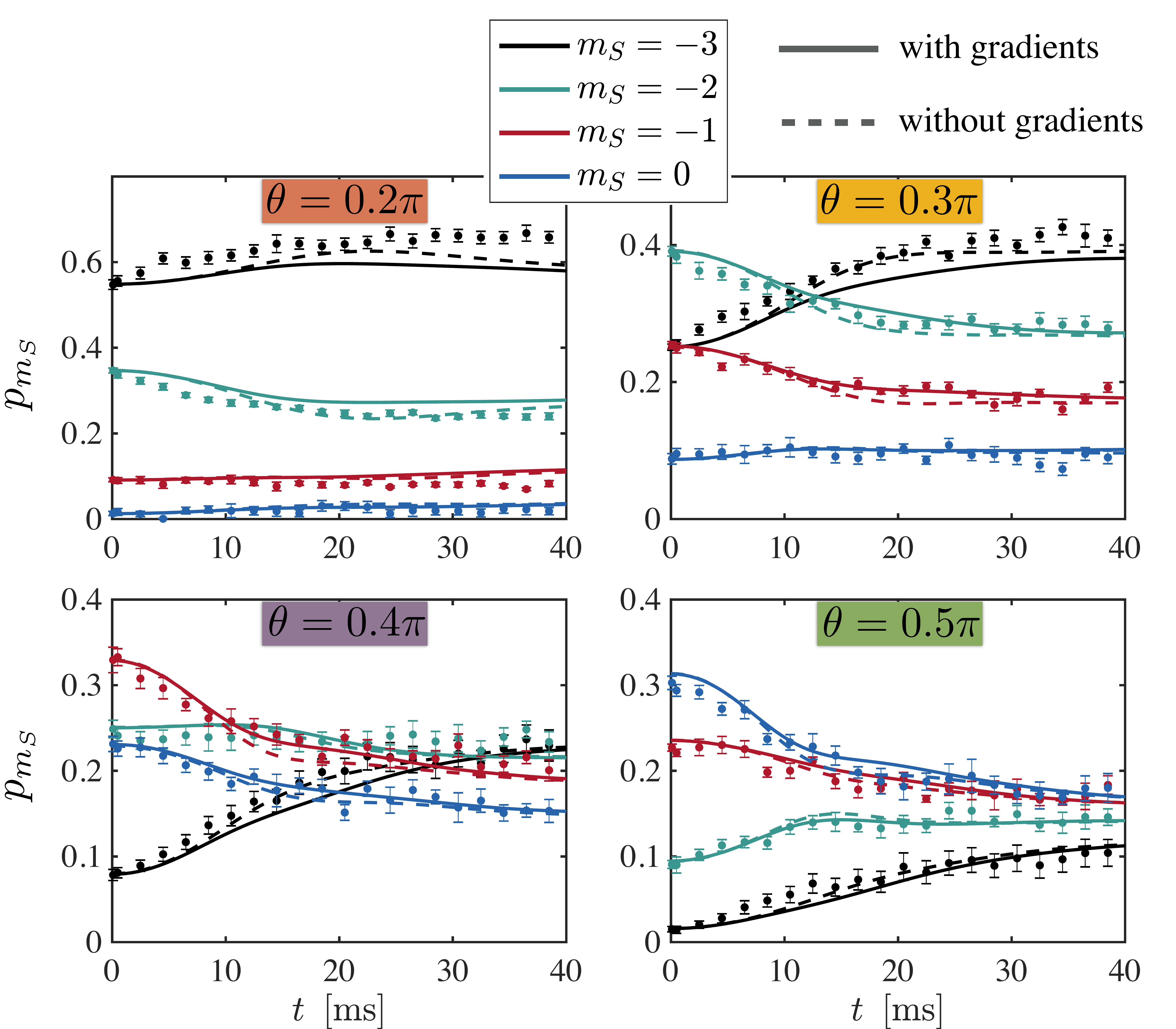}
\caption{{{\em The effect of the magnetic field gradient} is to slightly inhibit dynamics, after $\sim 10\,\text{ms}$. GDTWA simulations for the populations with identical parameters as in Fig.~2 of the main manuscript. Dashed lines show results for zero field gradient. The solid line is the usual result with the measured field gradient of $30\,\text{MHz/m}$ along the vertical axis.} }
\label{gradient}
\end{figure}

\begin{figure}
\centering
\includegraphics[width= 8.5 cm]{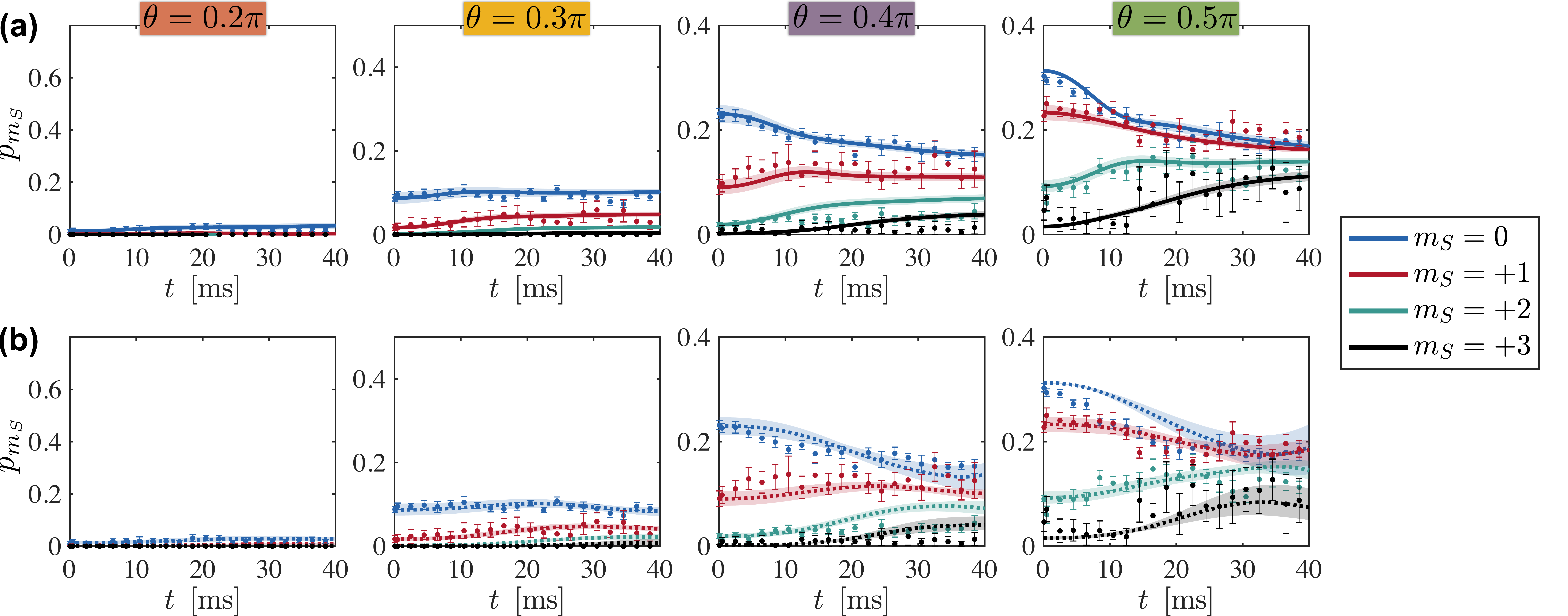}
\caption{{{\em Comparison between classical and quantum dynamics for upper spin levels.} Same panels as in Fig.~2 of the main manuscript, but for the spin levels $m=0,+1,+2,+3$.}  }
\label{mas}
\end{figure}

\subsection{Effect of none ideal preparation} The effect of defects due to imperfect preparation  is very small.
The presence of $10\%$ hole defects barely modifies the dynamics. See Fig~\ref{holes}.

\subsection{Effect of magnetic field gradients } We also investigated the effect of magnetic field gradients. This effect is very small at short times (as expected from perturbation theory, see Eq. (3) of the main text), and only slightly inhibits dynamics at longer times, as shown in Fig.\ref{gradient}

\subsection{Dynamics of  upper Zeeman sublevels not shown in main text } These are shown in Fig.~\ref{mas}.
They were omitted in the main text to make the figure less crowded.

\end{document}